\documentclass[number,preprint,12pt]{elsarticle}

\usepackage{subeqn}
\usepackage{graphicx}

\usepackage{amssymb}

\usepackage{lineno}

\biboptions{square,comma}

\newcommand{\bc}{{k_B}}

\newcommand{\Pe}{{\mbox{\upshape{Pe}}\:}} 
\newcommand{\tinyPe}{{\mbox{\scriptsize{\upshape{Pe}}}\:}} 
\newcommand{\pd}[2]{\frac{\partial #1}{\partial #2}}

\def\mathbi#1{\textbf{\em #1}}
\DeclareMathSymbol{\Delta}{\mathalpha}{letters}{"01}

\graphicspath{{./../figures/}}

\journal{Journal of Colloid and Interface Science}

\begin{document}

\begin{frontmatter}



\title{Partition-Induced Vector Chromatography in Microfluidic Devices}


\author{Jorge A. Bernate}
\ead{jbernate@jhu.edu}
\author{German Drazer\corref{cor1}\fnref{fn}}
\ead{drazer@jhu.edu}
\ead[url]{http://microfluidics.jhu.edu/}

\cortext[cor1]{Corresponding author}
\fntext[fn]{Tlf: (001) 410-516-0170, Fax: (001) 410-516-5510}
\address{Department of Chemical and Biomolecular Engineering, Johns Hopkins University, 221 Maryland Hall 3400 North Charles Street, Baltimore, Maryland 21218, USA}

\begin{abstract}
The transport of Brownian particles in a slit geometry in the presence of an arbitrary two-dimensional periodic energy landscape and driven by an external force or convected by a flow field  is investigated by means of macrotransport theory. Analytical expressions for the probability distribution and the average migration angle of the particles are obtained under the Fick-Jackobs approximation. The migration angle is shown to differ from the orientation angle of the driving field and to strongly depend on the physical properties of the suspended species, thus providing the basis for vector chormatography, in which different species move in different directions and can be continuously fractionated. The potential of microfluidic devices as a platform for partition-induced vector chromatography is demonstrated by considering the particular case of a piece-wise constant, periodic potential that, in equilibrium, induces the spontaneous partition of different species into high and low concentration stripes, and which can be easily fabricated by patterning physically or chemically one of the surfaces of a channel. The feasibility to separate different particles of the same and different size is shown for systems in which partition is induced via 1g-gravity and Van der Waals interactions in physically and chemically patterned channels, respectively. 
\end{abstract}


\end{frontmatter}
\section{Introduction}

Separation of the different constituents of a complex sample has long been of paramount relevance in many fields of engineering and science. The ubiquity of particle separation, for example, has led to the development of a vast number of different techniques~\citep{UnifiedSepSci:1991}. The manipulation of chemical and biological species at the micro and nanoscale  has also received special attention in the ongoing effort toward process miniaturization, with nearly every advance in our understanding of the effects that are dominant at these scales leading to the innovation of different separation devices as surveyed in recent reviews~\citep{Pamme:2007,Kulrattanarak:2008,Kovarik:2009}. An ideal technique would be able to discriminate particles based on small differences on any of a wide range of physicochemical properties, thus allowing the fractionation of intricate mixtures. In addition, continuous operation is generally preferred over a batch process because of the ease of operation and higher yields. To facilitate operation and portability it is also desirable that the device be autonomous, thus eliminating the need for external components. In contrast, most techniques discriminate particles based on a single physicochemical property,  are inherently batch-processes, or require external fields to drive separative displacement. 

One of the most versatile separation methods is Field-flow fractionation $-$FFF$-$~\citep{Giddings:1993}, a family of techniques in which a variety of force fields can be used to induce separation based on different physicochemical properties, with subtechniques including gravitational, sedimentation, flow, thermal, and electrical FFF~\citep{Eijkel:2006,Kowalkowski:2006}. In FFF, particles are typically transported by a parabolic flow between two parallel plates and, at the same time, externally forced in the direction perpendicular to the flow toward one of the walls. The velocity of the particles is thus governed by their equilibrium particle-wall separation, with particles in closer proximity to the wall moving slower, which results in the separation of the sample into bands of particles moving in the direction of the flow. Dielectrophorectic methods~\citep{Gascoyne:2002,Hughes:2002} are also versatile since differences in a broad spectrum of properties result in different dielectric signatures allowing, similarly to FFF, the separation of multifarious mixtures.

Despite the breadth of FFF techniques, they share the common drawback of being fundamentally a batch process. A technique as wide-ranging as FFF that allows for continuous operation is Split-Flow Lateral-Transport Thin $-$SPLITT$-$ fractionation~\citep{Giddings:1985}. In this technique, particles are streamed near one of the walls of the SPLITT channel and, at the same time, an external force transports  particles selectively in the direction perpendicular to the carrier flow into different streams which can then be continuously collected. On the other hand, SPLITT only allows bi-modal separation in a single stage. Other methods also allowing continuous operation have been developed by acoustic and optical means. Acoustic forces have been used to separate particles according to their density and compressibility into two streams at the pressure node and antinode created by a standing wave between two parallel walls~\citep{Evander:2008}. Optical forces have allowed the sorting of nano and microparticles of different materials and sizes, as reviewed by Jon{\'a\v{s}} and Zem{\'a}nek~\citep{Jonas:2008}. These continuous methods, however, require the integration with external components.

Continuous separation methods that do not require the presence of external fields include pinched flow fractionation $-$PFF$-$~\citep{Yamada:2004,Takagi:2005,Yuushi:2006}, hydrodynamic filtration $-$HF$-$~\citep{Yamada:2006}, Hydrophoresis~\citep{Sungyoung:2007,Sungyoung:LabChip2009,Sungyoung:AnalChem2009}, and inertial focusing~\citep{DiCarlo:2007,Bhagat:2008,Park:2009}. In PFF particles exit a pinched flow region in different stream lines when they are initially aligned to one of the walls of the constriction. Similarly, in HF particles are  first aligned along the walls of a main channel and then sequentially collected according to their size by controlling the flow rate in side channels. Hydrophoresis, on the other hand, does not require precise flow focusing of the particles. Hydrophoretic separation is achieved using an array of obstacles to induce a pressure field  responsible for the selective displacement of particles of different size. More recently, inertial lift forces and Dean flows in curved conduits have been used to stream particles in precise locations within a flow channel. However, hydrophoresis and inertial focusing methods are difficult to implement for the simultaneous collection of particles in polydispersed samples. 

Two dimensional $-$2D$-$ separation is another approach that allows for continuous fractionation. The spatial resolution required for continuous operation in 2D methods is achieved by a combination of transport in one direction with selective displacement in the perpendicular direction and results in greater resolving power compared to 1D techniques~\citep{Giddings:1984}. These methods fall in the category {\emph{vector chromatography}}~\citep{Dorfman:2001,Dorfman:2002}$-$VC$-$ tecniques, where the fractionation relies on differences in the average direction in which species being separated move. An extension of FFF, for example, combines radial and tangential carrier flows between two parallel disks to allow for the continuous collection of separated bands in different locations of the fractionation cell~\citep{Vastamaki:2005}. Other examples of 2D fractionation use a  flow field and an external force applied in the direction perpendicular to the flow as the selective displacement. In magnetophoresis~\citep{Pamme:2004}, for instance, particles are sorted according their magnetic susceptibility and size. 

A promising trend in VC exploits the interactions between the species being separated and features intentionally designed in the separation media to drive separative displacement. For example, deterministic lateral displacement~\citep{Huang:2004,Davis:2006,Mao:2009,Li:2007,MortonLabChip:2008,MortonPNAS:2008} $-$DLD$-$ is carried out in devices patterned with a 2D sieving matrix in which particles of different sizes move at different angles, thus enabling the simultaneous fractionation and collection of multiple components in a polydispersed sample with very high resolution. However, the underlying mechanism leading to separation in these sieving devices is not completely clear, which has led so far to ad hoc designs that are difficult to optimize~\citep{Herrmann:2009,Drazer:2009,Balvin:2009}. In this context, Dorfman and Brenner~\citep{Dorfman:2001} considered the illustrative case of a periodic system consisting of repeating layers of two inmiscible fluids. They showed that differences in  the partition ration of the particles $-$between the two phases$-$, due to differences in affinity between the particles and each of the layers of fluid, results in vector chromatography  when the particles are animated by a constant external force. Unfortunately, as noted by the authors~\citep{Dorfman:2001}, such a two-fluid layer system cannot be easily implemented in practice.   

In this study we demonstrate the potential of planar microfluidic devices as a platform to achieve {\emph{partition-induced vector chromatography}} $-$PIVC$-$, a technique that combines the versatility of FFF with the multiple advantages of continuous 2D vector separation. The high surface-to-volume ratio characteristic of microfluidic devices makes it possible to induce partition by means of surface interactions. In fact, it has been shown that the spontaneous partition of different species can be controlled by means of the energy landscape created by a chemical or physical pattern on one of the surfaces of a channel~\citep{Wu:2006,Bahukudumbi:2007}. In particular, we consider cases in which partition is induced via Van der Waals forces and 1-g gravity, both approaches resulting in autonomous devices. In addition, external fields could also be used to cause or enhance partition. We show that partition results in diffusive fluxes that are responsible for the migration of particles at angles different from the orientation angle of the driving field.

The content of this work is structured as follows. First, the equation governing the probability density is derived for a spatially periodic potential following the macrotransport paradigm~\citep{MacroTransport}. In the subsequent sections, this governing equation is solved under the Fick-Jacobs $-$FJ$-$ approximation for the cases in which the particle is animated either by a constant external force or by a fluid flow. Closed-form solutions for the particle trajectory angle in a slit geometry are obtained for an arbitrary two-dimensional potential. These general solutions are evaluated to obtain analytical expressions for a piece-wise constant potential. Then, it is shown how such a potential can be achieved by patterning chemically or physically  one of the surfaces of a microfluidic device with an array of rectangular stripes. Specifically, we consider the case in which partition between the stripes is induced by differences in the potential energy resulting from 1-g gravity, Van der Waals, and electrostatic forces. Finally, we look at experimentally accessible systems and validate the results obtained with the FJ approximation with Brownian dynamics simulations to show that particles exhibiting different partition ratios can be effectively fractionated via vector chromatography.

\section{Particle transport in patterned microfluidic devices: Macrotransport theory}
\begin{figure}[htbp]
	\begin{center}
		\begin{tabular}{|c|}
			\hline
			 \parbox[][][c]{7.5cm}{
	          		\includegraphics[width=7.5cm]{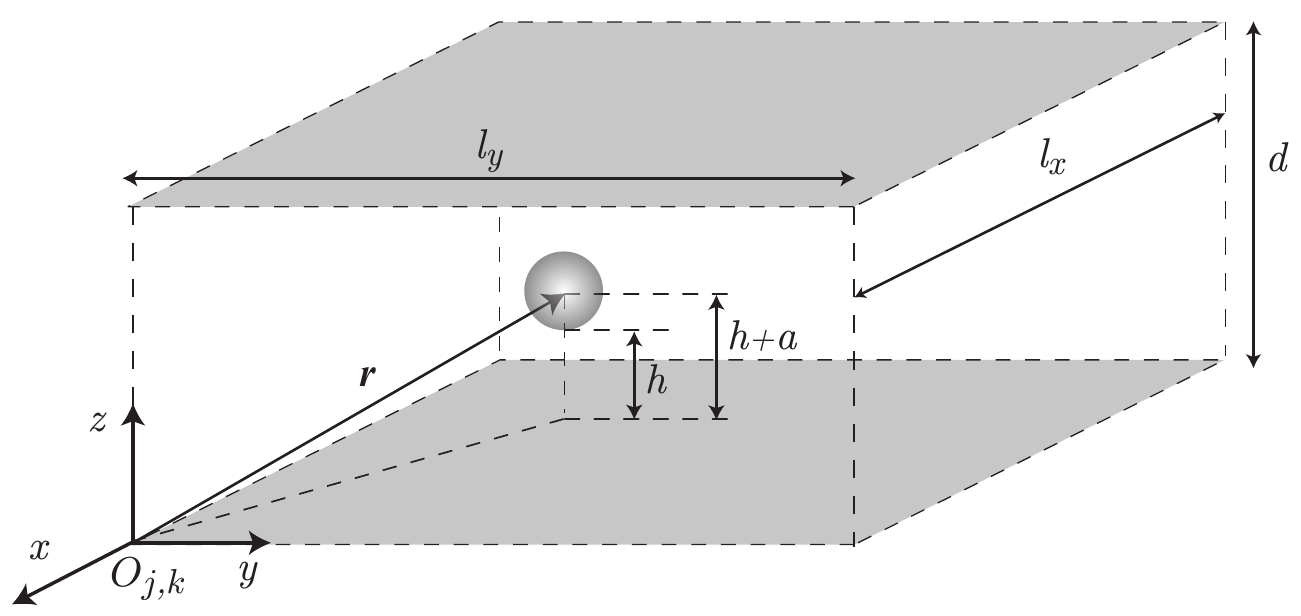}}\\ 
			\hline			
		\end{tabular}
	\end{center}
\caption[Unit cell]{The particle is suspended between two parallel walls separated a distance d $-$slit geometry$-$ and the physicochemical properties of the impermeable walls are such that the system is periodic in the $x$ and $y$ directions.}
\label{Fig:slit_periodicity}
\end{figure}

In this section we derive, by means of macrotransport theory~\citep{MacroTransport}, the macroscopic equations governing the transport of suspended particles in the periodic system shown in Fig.~\ref{Fig:slit_periodicity}. The particles are confined between two infinite parallel walls separated by a distance $d$. We shall assume that the physicochemical properties of the walls of the channel render the system  $l_x$- and $l_y$-periodic in the $x$ and $y$ directions, respectively. 
Let $P({\mathbi{R}},t|\mathbi{R}_0)$ be the conditional probability density of finding the particle center, modeled as a stochastic variable, at location ${\mathbi{R}} = \left(\mathrm{x},\mathrm{y},z\right)$ at time $t$, given the initial position ${\mathbi{R}}_0 = \left(\mathrm{x}_0,\mathrm{y}_0,z_0\right)$ at time $t=0$. This conditional probability density is governed by the following convection diffusion equation,
\begin{equation}
\pd{P}{t} +  \nabla \cdot {\mathbi{J}} = \delta({\mathbi{R}}-{{\mathbi{R}}_0})\delta(t),
\label{Eq:microP}
\end{equation}
where the right-hand side represents an instantaneous unit impulse at ${\mathbi{R}}_0$ at time $t=0$. The flux $\mathbi{J}$ is composed of a convective and a diffusive contribution, the latter assumed to have Fickian form,
\begin{equation}
{\mathbi{J}}= {\mathbi{U}} P - {\mathbi{D}}\cdot\nabla P,
\label{Eq:J}
\end{equation}
where $\mathbi{U}$ is the net convective velocity of the particle due to both flow and applied forces, and $\mathbi{D}$ is the diffusion tensor. Due to the periodicity of the system, $P({\mathbi{R}},t|{\mathbi{R}}_0)$ can be written in the equivalent functional form
$P({\mathbi{R}}_{n,m},{\mathbi{r}},t|{\mathbi{R}}_{n_0,m_0},{\mathbi{r}}_0)$
where 
${\mathbi{R}}_{j,k} = j \, l_x \; {\bf{i}}_x + k \, l_y \;{\bf{i}}_y$   represents the location of the $(j,k)$th unit cell with respect to some arbitrary origin, and $\mathbi{r} = (x,y,z)$ is the intracell position vector. 
In addition, both ${\mathbi{U}}({\mathbi{r}})$ and ${\mathbi{D}}({\mathbi{r}})$ are periodic and dependent on the intracell position ${\mathbi{r}}$ only. Therefore, the probability distribution function $P$ does not depend on $\mathbi{R}_{n,m}$ and $\mathbi{R}_{n_0,m_0}$ independently but on their difference, $P(\mathbi{R}_{n,m}-\mathbi{R}_{n_0,m_0},{\mathbi{r}},t|{\bf{r}}_0)$. Then, the zeroth order moment of the distribution,
\begin{equation}
P_0(\mathbi{r},t|\mathbi{r}_0) = \sum_{n,m} P(\mathbi{R}_{n,m}-\mathbi{R}_{n_0,m_0},{\mathbi{r}},t|{\mathbi{r}}_0),
\end{equation}
represents the conditional probability density that the particle will be found in the local, intracellular position ${\mathbi{r}}$ at time $t$, given that it was introduced at the location ${\mathbi{r}}_0$ at time $t=0$. $P_0({\mathbi{r}},t|{{\mathbi{r}}_0})$ satisfies the equation
\begin{equation}
\pd{P_0}{t}+\nabla \cdot {{\mathbi{J}}_0} = \delta({\mathbi{r}}-{{\mathbi{r}}_0})\delta(t),
\end{equation}
where the flux ${{\mathbi{J}}_0}$ follows Eq.~(\ref{Eq:J}). This governing equation for the probability distribution corresponds to the Smoluchosky equation describing the stochastic motion of a particle in the limit of negligible inertia~\citep{FokkerPlanck}. The steady state solution, 
\begin{equation}
P^{\infty}_0({\bf{r}}) = \lim_{t \to \infty} P({\mathbi{r}},t|{{\mathbi{r}}_0}),
\end{equation}
is then governed by the equation
\begin{equation}
\nabla \cdot {\mathbi{J}}^{\infty}_0 = 0,
\label{Eq:deldotJinfty_0}
\end{equation}
with periodic boundary conditions,
\begin{subequations} 
\begin{equation}
P^{\infty}_0(x,y,z) = P^{\infty}_0(x+l_x,y,z),
\end{equation}
\begin{equation}
P^{\infty}_0(x,y,z) = P^{\infty}_0(x,y+l_y,z),
\end{equation}
\label{periodicBCs} 
\end{subequations}
and the no-flux condition enforced on the impermeable walls confining the particle in the $z$-direction. Finally, the particle is bound to be somewhere within the volume $\tau$ of the unit cell, which implies the standard normalization condition 
\begin{equation}
\int_{\tau}P^{\infty}_0({\mathbi{r}})d^3 r = 1.
\label{normalizationC}
\end{equation}

The asymptotic flux of probability density can be written in our case as
\begin{equation}
{\mathbi{J}}^{\infty}_0 = {\mathbi{u}}P^{\infty}_0 + {\mathbi{M}} \cdot {\mbox{\mathbi{\itshape{F}}}} P^{\infty}_0 -{\mathbi{M}}\cdot \nabla V P^{\infty}_0 -{\mathbi{D}}\cdot \nabla P^{\infty}_0,
\label{Eq:Jinfty_0}
\end{equation}
where $\mathbi{u}({\mathbi{r}})$ is the velocity of the particle when convected by a flow field, ${\mathbi{F}}$ is a constant external force acting on the particle, $V({\mathbi{r}})$ is the potential energy landscape in which the particle is moving, and ${\mathbi{M}}({\mathbi{r}})$ is the mobility tensor, and all quantities are assumed to be periodic. 
For the geometry considered here, it is convenient to write the mobility and diffusion tensors as
\begin{subequations}
\begin{equation}
{\mathbi{M}} = {\mathbi{i}}_z{\mathbi{i}}_z M_{\bot}  + ({\mathbi{I}}-{\mathbi{i}}_z{\mathbi{i}}_z)M_{||},
\end{equation}
\begin{equation}
{\mathbi{D}} = {\mathbi{i}}_z{\mathbi{i}}_z D_{\bot}  + ({\mathbi{I}}-{\mathbi{i}}_z{\mathbi{i}}_z)D_{||}.
\end{equation}
\end{subequations}
The mobilities and diffusivities follow the Stokes-Einstein relation, $D_{\bot} = \bc T M_{\bot}$ and $D_{||} = \bc T M_{||}$, where $\bc$ is the  Boltzmann constant and $T$ the absolute temperature. 

Knowledge of $P^{\infty}_0$ yields the mean particle velocity from the quadrature
\begin{equation}
{\bar{\mathbi{U}}}^* = \int_{\tau}{\mathbi{J}}^{\infty}_0 d^3 r.
\end{equation}
The orientation angle of ${\bar{\mathbi{U}}}^*$ is the relevant parameter in vector chromatography, termed hereafter the chromatrographic trajectory angle, more simply the trajectory angle \citep{Dorfman:2001}, or the migration angle. In the following sections we derive the trajectory angle under the Fick-Jacobs approximation for the cases in which the particle is driven either by an externally applied force or by a flow field.
 

\subsection{Non-equilibrium transport: The Fick-Jacobs approximation}

The three-dimensional problem posed in the preceding section can be solved numerically for a particular geometry and driving fields. In this section, we focus on essentially 2D problems that capture the fundamental mechanisms leading to PIVC, and which are amenable analytically. Specifically, we consider the $x$-invariant case whereby ${\mathbi{u}}(\mathbi{r})$,  $M(\mathbi{r})$, $V(\mathbi{r})$ and $P^{\infty}_0(\mathbi{r})$ depend on $y$ and $z$ only, and the periodicity in the $x$-direction is automatically satisfied. In addition, we study 2D driving fields in which the external force and the fluid-flow velocity far from the particle are in the $x$-$y$ plane. Under these conditions the asymptotic flux, Eq.~(\ref{Eq:Jinfty_0}), takes the form
\begin{eqnarray}
{\mathbi{J}}^{\infty}_0 &=&  \left( u_x P^{\infty}_0 + M_{||} F_x P^{\infty}_0 \right) {\mathbi{i}}_x  \nonumber\\
                   &+&  \left( u_y P^{\infty}_0 + M_{||}F_y P^{\infty}_0 - M_{||}\pd{V}{y}P^{\infty}_0 - \bc T M_{||} \pd{P^{\infty}_0}{y} \right){\mathbi{i}}_y\\
                    &+&  \left( - M_{\bot}\pd{V}{z}P^{\infty}_0 - \bc T M_{\bot} \pd{P^{\infty}_0}{z} \right){\mathbi{i}}_z, \nonumber
\end{eqnarray}
and Eq.~(\ref{Eq:deldotJinfty_0}), governing the asymptotic probability density, becomes
\begin{eqnarray}
0 &=&  \pd{}{y}\left(u_y P^{\infty}_0 + M_{||} F_y P^{\infty}_0 - M_{||}\pd{V}{y}P^{\infty}_0 - \bc T M_{||} \pd{P^{\infty}_0}{y} \right)  \nonumber  \\
 &+& \pd{}{z}\left( -M_{\bot}\pd{V}{z}P^{\infty}_0 - \bc T M_{\bot} \pd{P^{\infty}_0}{z}\right).
\label{GE}
\end{eqnarray}

A heuristic approximation first introduced by Jacobs~\citep{DiffusionProcesses,Zwanzig:1992} allows solving Eq.~(\ref{GE}) analytically assuming local equilibrium in the transverse direction. This approximation, termed by Zwanzig the Fick-Jacobs (FJ) approximation, has been shown to provide accurate results in narrow channels~\citep{Burada:2007,Laachi:2007,XinLi:2009}. Under this local equilibrium assumption, and given that the system is $x$-invariant, the asymptotic probability distribution conditioned to a given intracelular position $y$ can be written as
\begin{equation}
\rho(x,z|y) \approx \rho(x,z|y)_{eq} = \left.e^{-V(y,z)/\bc T}\right/ I(y),
\label{rhoEq}
\end{equation} 
where 
\begin{equation}
I(y) = \int \int{e^{-V(y,z)/\bc T}}dz dx.
\end{equation}
Therefore, the position of the particle in $y$ suffices to determine the distribution, which takes the form
\begin{equation}
P^{\infty}_{0}(\mathbi{r}) = {\cal{P}}(y) \rho(x,z|y) \approx {\cal{P}}(y) \rho(x,z|y)_{eq} =  \left.{\cal{P}}(y)e^{-V(y,z)/\bc T}\right/I(y),
\end{equation}
where 
\begin{equation}
{\cal{P}}(y) = \int \int P^{\infty}_{0}(\mathbi{r}) dz dx
\end{equation}
is the marginal probability density. The cases in which the particle motion is driven either by a constant external force or by a fluid flow are considered under the FJ approximation in what follows. 

\subsubsection{External force} 
When the particle is animated solely by a constant external force, Eq.~(\ref{GE}) reads 
\begin{eqnarray}
0 &=& \pd{}{y}\left(M_{||}F_y P^{\infty}_0 - M_{||}\pd{V}{y}P^{\infty}_0 - \bc T M_{||} \pd{P^{\infty}_0}{y} \right) \nonumber\\
  &+& \pd{}{z}\left( -M_{\bot}\pd{V}{z}P^{\infty}_0 - \bc T M_{\bot} \pd{P^{\infty}_0}{z}\right). 
\label{GE_F}  
\end{eqnarray} 
Nondimensionalizing with the variable changes $x \to x/l_x$, $y \to y/l_y$, $z \to z/d$, $M_{||} \to M_{||}/ M_{\infty}$ and $M_{\bot} \to M_{\bot}/ M_{\infty}$ $-$where $M_{\infty} = 6 \pi \mu a$ being the mobility of an unbounded particle of radius $a$ in a fluid of viscosity $\mu$$-$, $V \to V/\bc T$, and $P^{\infty}_0 \to \tau P^{\infty}_0$, substituting the FJ approximation for $P^{\infty}_{0}$, integrating in $x$ and $z$, and rearranging the terms yields
\begin{eqnarray}
0 &=& \frac{d}{dy} \int \int J^{\infty}_{0_y}(\mathbi{r})dz dx \nonumber \\
  &=& \frac{d}{dy} \left\{ \left[\int \int M_{||}(y,z)e^{-V(y,z)}dz dx\right] \left\{ \Pe \frac{{\cal{P}}(y)}{I(y)} - \frac{d}{dy}\left[\frac{{\cal{P}}(y)}{I(y)}\right] \right\}\right\} 
\end{eqnarray}
where the only dimensionless parameter is the P\'eclet number, $\Pe = F_y l_y /\bc T$. The  total dimensionless flux in the $y$-direction, $J$, is clearly constant and can be written as
\begin{equation}
J = \int \int J^{\infty}_{0_y}(\mathbi{r})dz dx = I(y)\bar{M}_{||}(y) \left[ \: \Pe \tilde{\cal{P}}(y) -  \frac{d\tilde{\cal{P}}(y)}{dy}\right],
\label{J_force} 
\end{equation}
where 
\begin{equation}
\bar{M}_{||}(y) = \left.\int \int M_{||}(y,z)e^{-V(y,z)}dz dx \right/I(y),
\label{Mbar}
\end{equation}
is the local average of the mobility, and ${\tilde{\cal{P}}}(y) = \left.{\cal{P}}(y)\right/I(y)$.
The general solution for the marginal density is given by 
\begin{equation}
{\cal{P}}(y) = e^{\tinyPe y}\left[  -J\int^y_0 \frac{dy'}{e^{\tinyPe y'}I(y')\bar{M}_{||}(y')} + N\right]I(y).
\end{equation}
The flux $J$ and the constant $N$ are obtained by solving simultaneously the periodicity and normalization conditions:
\begin{subequations}
\begin{equation}
N = \frac{1}{\Sigma}\int^{1}_{0} \frac{dy'}{e^{\tinyPe y'}I(y')\bar{M}_{||}(y')},
\end{equation}
\begin{equation}
J = \frac{1}{\Sigma}\left( 1-e^{-\tinyPe}\right),
\label{Eq:J2}
\end{equation}
\begin{equation}
\Sigma = \int^1_0 dy e^{\tinyPe y}I(y)\int^{y+1}_y \frac{dy'}{e^{\tinyPe y'} I(y')\bar{M}_{||}(y')}. 
\end{equation}
\label{Eq:N_J_Sigma_Force}
\end{subequations}
The total flux in the $z$ direction is identically equal to zero. Thus,
\begin{equation}
{\bar{{\mathbi{U}}}}^* = U^*_x \: {\mathbi{i}}_x +  U^*_y \: {\mathbi{i}}_y = \int_{\tau}J^{\infty}_{0_x} d^3 r \: {\mathbi{i}}_x + \int_{\tau}J^{\infty}_{0_y} d^3 r \: {\mathbi{i}}_y
\label{Eq:Ustar}
\end{equation}
Employing Eqs.~(\ref{Eq:N_J_Sigma_Force}) and~(\ref{Eq:Ustar}), we obtain one of our main results, a chromatographic trajectory angle that depends in general on the properties of the particle through the energy landscape $V$ and the external force $F$:
\begin{equation}
\tan\theta^* = \frac{U^*_x }{U^*_y } = \tan \theta_F  \frac{\Pe}{1-e^{-\tinyPe}} \int^1_0 dy e^{\tinyPe y} I(y)\bar{M}_{||}(y) \int^{y+1}_y \frac{dy'}{I(y')\bar{M}_{||}(y')e^{\tinyPe y'}},
\label{Eq:theta_F_arbitrary}
\end{equation}
where $\theta_F = \arctan(F_x/F_y)$ is the orientation angle of the external force.

\subsubsection{Entrainment in a fluid flow}
When the particle is convected by a fluid flow, Eq.~(\ref{GE}) reduces to 
\begin{eqnarray}
0 &=& \pd{}{y}\left(u_y P^{\infty}_0 - M_{||}\pd{V}{y}P^{\infty}_0 - \bc T M_{||} \pd{P^{\infty}_0}{y} \right) \nonumber\\
  &+& \pd{}{z}\left( -M_{\bot}\pd{V}{z}P^{\infty}_0 - \bc T M_{\bot} \pd{P^{\infty}_0}{z}\right) 
\label{GE_u} 
\end{eqnarray}
Nondimensionalizing with the same change of variables as those used for Eq.~(\ref{GE_F}) in addition to $u_y \to u_y/U$, where $U$ is the average velocity of the particle-free flow, substituting the FJ approximation for $P^{\infty}_0$, integrating in $x$ and $z$, and rearranging the resulting terms we obtain the $-$constant$-$ total flux in the $y$ direction 
\begin{equation}
J = \int \int J^{\infty}_{0_y}(\mathbi{r})dz dx =  I(y) \bar{M}_{||}(y) \left[ \Pe \frac{\bar{u}_{y}(y)}{\bar{M}_{||}(y)}\tilde{\cal{P}}(y) -  \frac{d\tilde{\cal{P}}(y)}{dy}\right].
\label{pbar_flow}
\end{equation}
In this case, the P\'eclet number is $ \Pe = U l_y/\bc T M_{\infty}$ and $\bar{u}_{y}(y)$, analogously to $\bar{M}_{||}(y)$, is the local average of the velocity $u_y$. We can then define a local effective force as the ratio of the local average velocity to the local average mobility,
\begin{equation}
\phi(y) = \frac{\bar{u}_{y}(y)}{\bar{M}_{||}(y)}.
\end{equation}
Then, the general solution for the marginal probability distribution under the FJ approximation can be expressed as
\begin{equation}
{\cal{P}}(y)= e^{\tinyPe \Phi(y)}\left[ -J\int^y_0 \frac{dy'}{e^{\tinyPe \Phi(y')}I(y')\bar{M}_{||}(y')} + N\right]e^{-V(y,z)},
\end{equation}
where 
\begin{equation}
\Phi(y) = \int^y_0 \phi(y')dy'.
\end{equation}
The flux J and the constant $N$ are obtained, as in the external force case, by solving simultaneously the periodicity and normalization conditions:
\begin{subequations}
\begin{equation}
N = \frac{1}{\Sigma}\int^{1}_{0} \frac{dy'}{e^{\tinyPe \Phi(y')}I(y')\bar{M}_{||}(y')},
\end{equation}
\begin{equation}
J = \frac{1}{\Sigma}\left[1-e^{-\tinyPe \Phi(1)}\right], 
\end{equation}
\begin{equation}
\Sigma = \int^1_0 dy e^{\tinyPe \Phi(y)}I(y)\int^{y+1}_y \frac{dy'}{e^{\tinyPe \Phi(y')} I(y')\bar{M}_{||}(y')}. 
\end{equation}
\end{subequations}
Lastly, the chromatographic trajectory angle is given by
\begin{equation}
\tan\theta^* = \frac{U^*_x }{U^*_y } = \tan \theta_f \frac{\Pe}{1-e^{-\tinyPe \Phi(1)}} \int^1_0 dy e^{\tinyPe \Phi(y)} I(y)\bar{M}_{||}(y) \int^{y+1}_y \frac{dy'}{e^{\tinyPe \Phi(y')}I(y')\bar{M}_{||}(y')},  
\label{Eq:theta_Flow_arbitrary}
\end{equation}
where $\theta_f$ is the orientation angle of the particle-free flow. Eqs.~(\ref{Eq:theta_F_arbitrary}) and~(\ref{Eq:theta_Flow_arbitrary}) are valid for an arbitrary two dimensional potential, under the FJ approximation. These expressions are used in the next section to obtain analytical results for a potential induced by surface modifications in a microfluidic device. 

\section{Vector chromatography in microfluidic devices: Stripe patterns}
\begin{figure}[htbp]
	\begin{center}
		\begin{tabular}{|c|}
			\hline
			 \parbox[][][c]{10 cm}{
	          		\includegraphics[width= 10 cm]{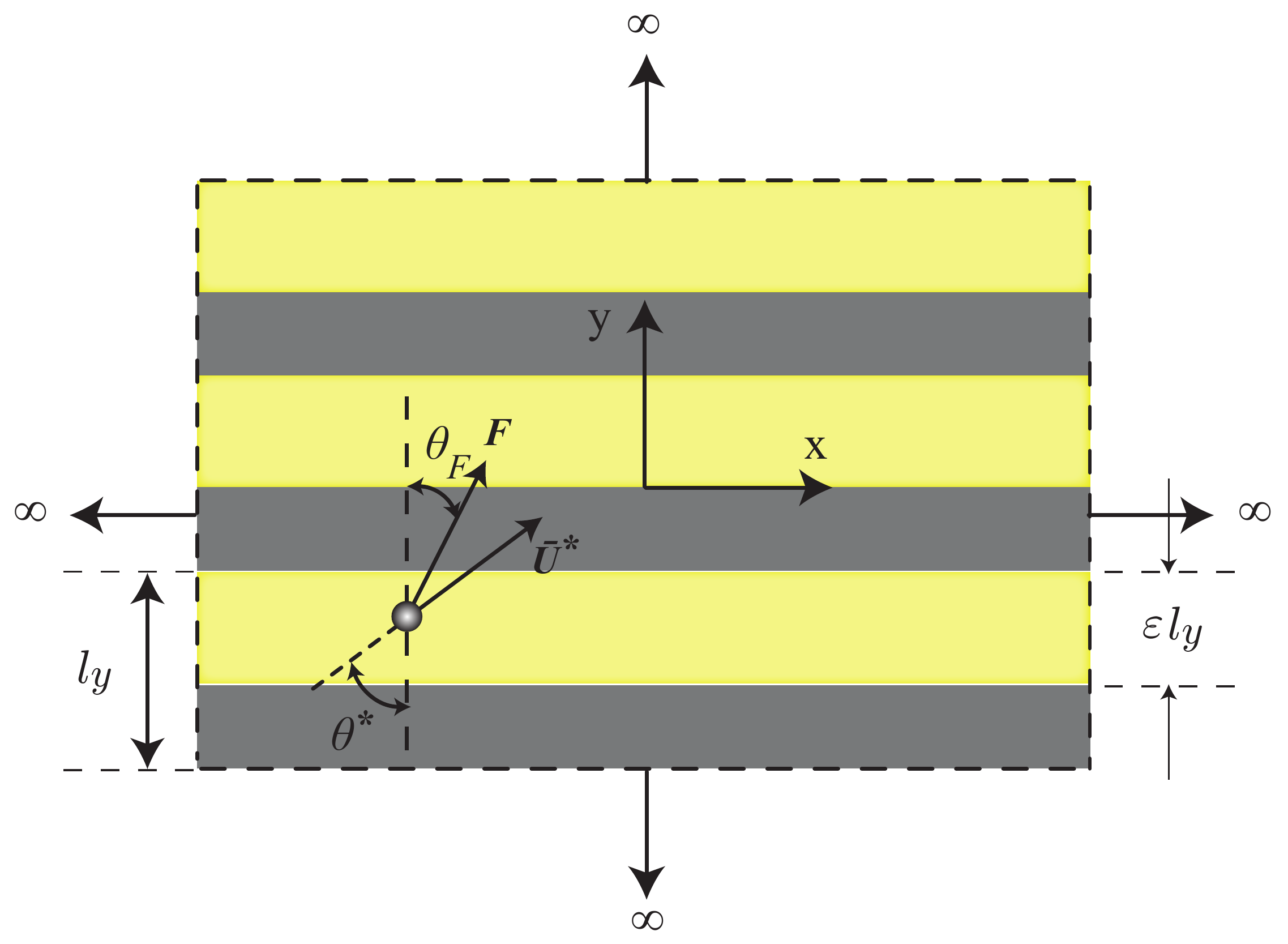}}\\ 
			 \hline		
		\end{tabular}
	\end{center}
\caption[Schematic]{Schematic view from the top of a particle driven by an external force. The particle is suspended over a pattern of rectangular stripes of widths $\epsilon$ and $1-\epsilon$ relative to the period length $l_y$.}
\label{Fig:Schematic}
\end{figure}

Using standard microfabrication techniques, a straight channel in a microfluidic device can be modified to yield a periodic system by patterning one of the surfaces of the channel with an array of rectangular stripes as shown in Fig.~\ref{Fig:Schematic}. We consider the case where the interfaces between the stripes are sharp, resulting for particles at close proximity to the wall, in approximately a piece-wise potential of the form
\begin{equation}
V(\mathbi{r}) = \left\{\begin{array}{c} {V}_{1}(z) \quad 0 < y < \epsilon   \\
                                        {V}_{2}(z)  \quad \epsilon < y < 1. \\
                           \end{array}
                    \right.
\label{Eq:piece_wise_potential}                    
\end{equation}


The migration angles can then be calculated by substituting this potential into the general expressions given by Eq.~(\ref{Eq:theta_F_arbitrary}) and~(\ref{Eq:theta_Flow_arbitrary}). It is worth mentioning that the transition potentials do not contribute to the trajectory angle whenever the transition regions are much narrower than the stripes, and therefore their explicit inclusion in Eq.~(\ref{Eq:piece_wise_potential}) is not necessary. This simple approximation provides analytical expressions that facilitate the understanding of the physics underlying the separation process, and shows that separation is feasible.

\subsection{Interaction potential}
In general, the potential on each stripe includes the Van der Waals and electrostatic forces between the particle and the surface and, for non-neutrally buoyant particles, it has a gravitational contribution as well,
\begin{equation}
V(h)=V_{edl}(h)+V_{vdw}(h)+V_{grav}(h),
\label{predicted_potential}
\end{equation}
where $h$ is the particle wall separation $-$see Fig.~\ref{Fig:slit_periodicity}$-$. The electrostatic double-layer potential for the case of a $z$-$z$ electrolyte has the form~\citep{Russel:1989}
\begin{equation}
V_{edl}(h) = a B \exp(-\kappa h)
\end{equation}
with
\begin{equation}
B = 64 \pi \epsilon_m \left(\frac{\bc T}{e}\right)^2  \tanh \left(\frac{e \psi_p}{4 \bc T} \right) \tanh \left( \frac{e \psi_s}{4 \bc T} \right),
\end{equation}
where $\epsilon_m$ is the dielectric permitivity of the medium, $e$ is the charge of an electron, $\psi_p$ and $\psi_s$ are the Stern potentials of the particle and the wall, respectively, and 
\begin{equation}
\kappa^{-1} = \left(\sum_i e^2 z^2 C_i N_A/\epsilon \bc T \right)^{-1/2}
\end{equation}
is the Debye length, where $C_i$ is the bulk electrolyte concentration of species $i$, $z$ is the valence of the ions, and $N_A$ is Avogadro's number.

\begin{figure}[htbp]
	\begin{center}
		\begin{tabular}{|c|c|}
			\hline
			$(a)$ & $(b)$ \\ 
			 \parbox[][][c]{6 cm}{
	          		\includegraphics[width= 6 cm]{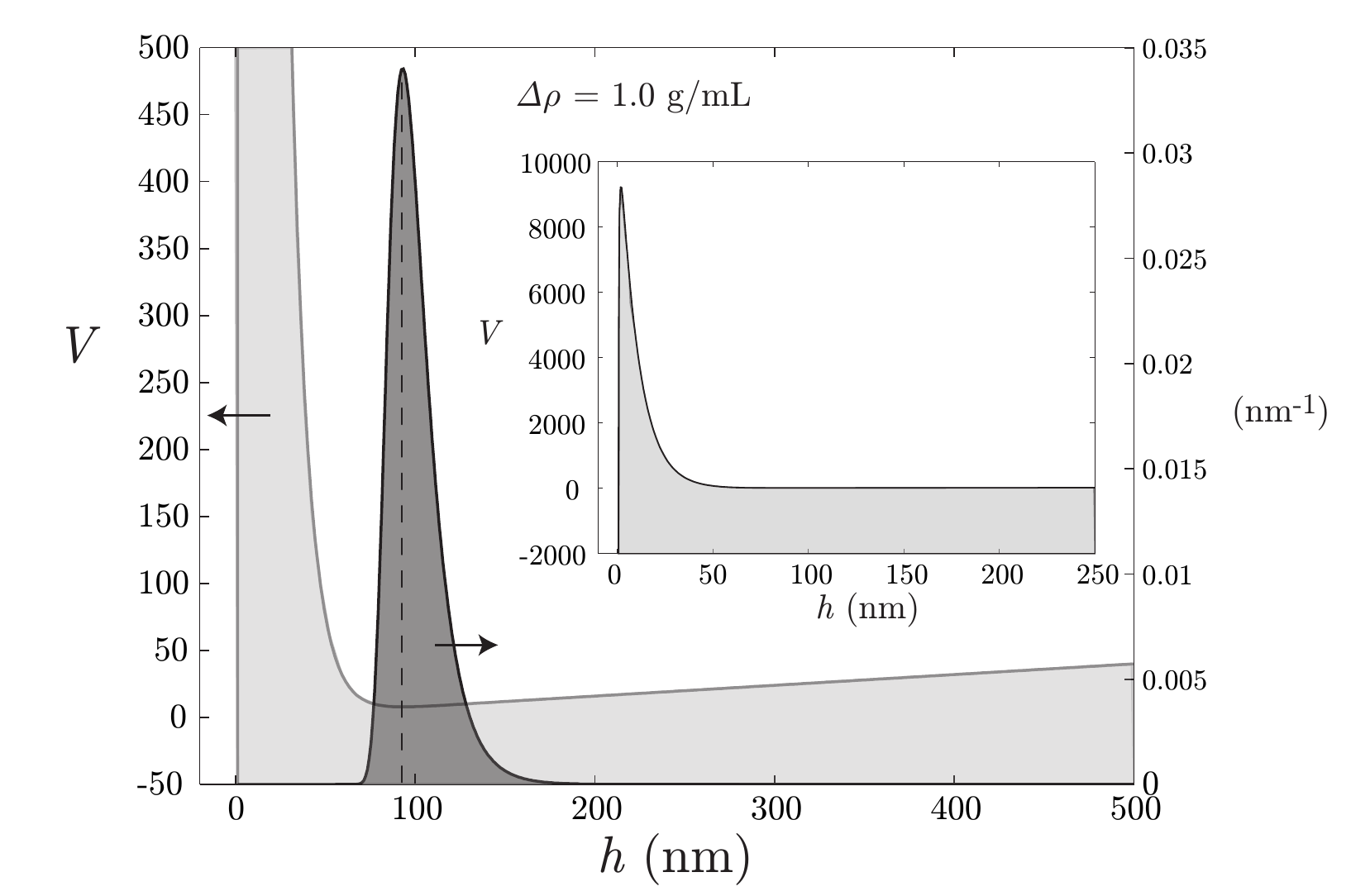}} &
	     \parbox[][][c]{6 cm}{
	               \includegraphics[width= 6 cm]{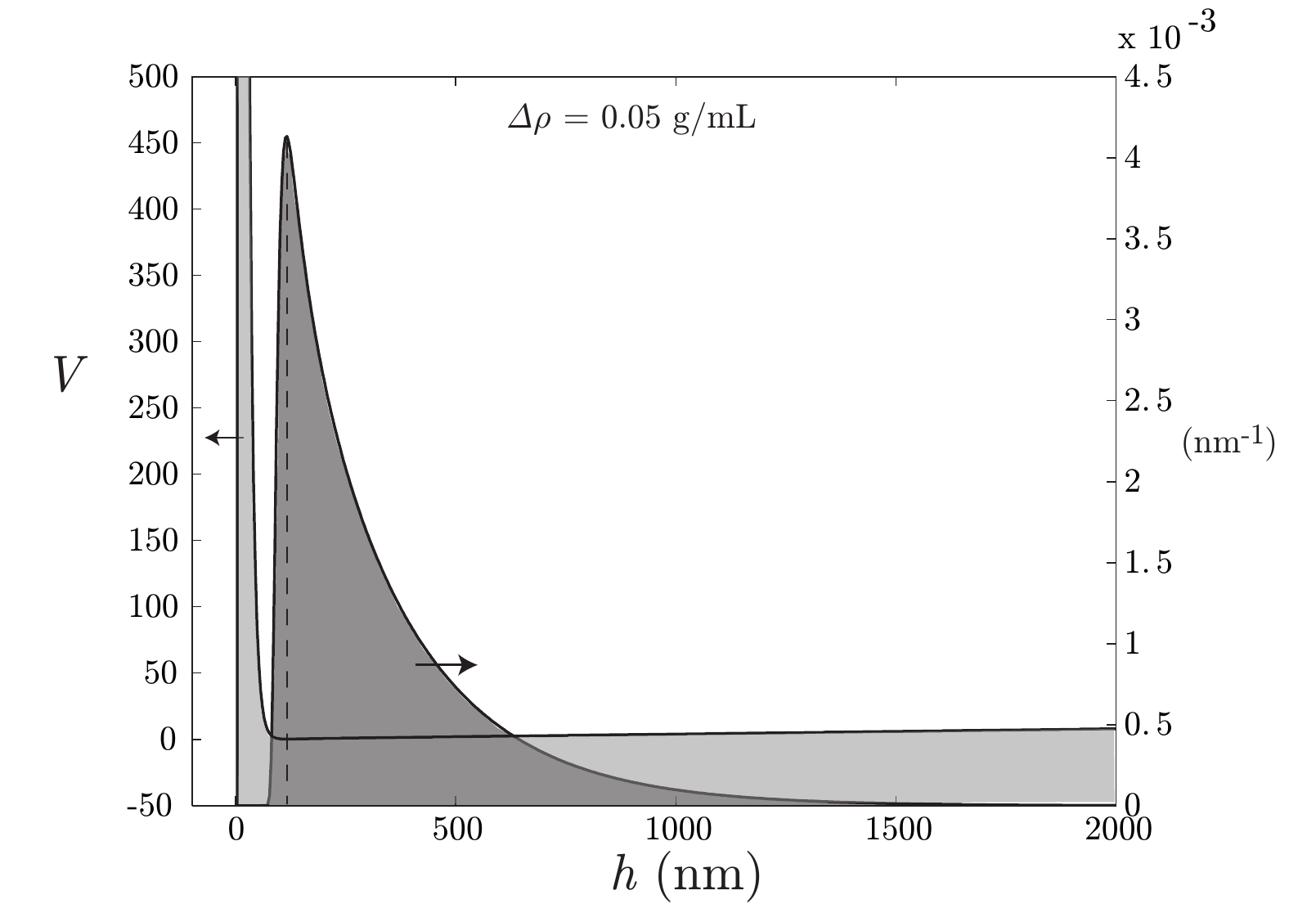}}\\ 
			 \hline				
			 \end{tabular}
	\end{center}
\caption[Potentials]{Total potential in $\bc T$ units $-$left axis, light shaded region$-$ and corresponding Boltzmann distribution $-$right axis, dark shaded region$-$ as a function of the particle-wall separation for $(a)$ a particle twice as dense and $(b)$ slightly denser than the aqueous medium. The inset in $(a)$ shows the repulsive energy barrier that has to be overcome to attain the primary minimum. The particle has a radius of $a = 2$ $\mu m$, the parameters of the electrostatic potential are $\psi_p = \psi_s = -60$ mV with an aqueous molarity of 1 mM  $-$1:1 electrolyte$-$, and the parameters of the Van der Waals potential are $A = 2$ nm,  $p = 2$. The surface and gravitational potentials vanish at an infinite separation and at the wall, respectively.} 
\label{Fig:potentials}
\end{figure}

\begin{figure}[htbp]
	\begin{center}
		\begin{tabular}{|c|}
			\hline
	     \parbox[][][c]{10 cm}{
	          		\includegraphics[width= 10 cm]{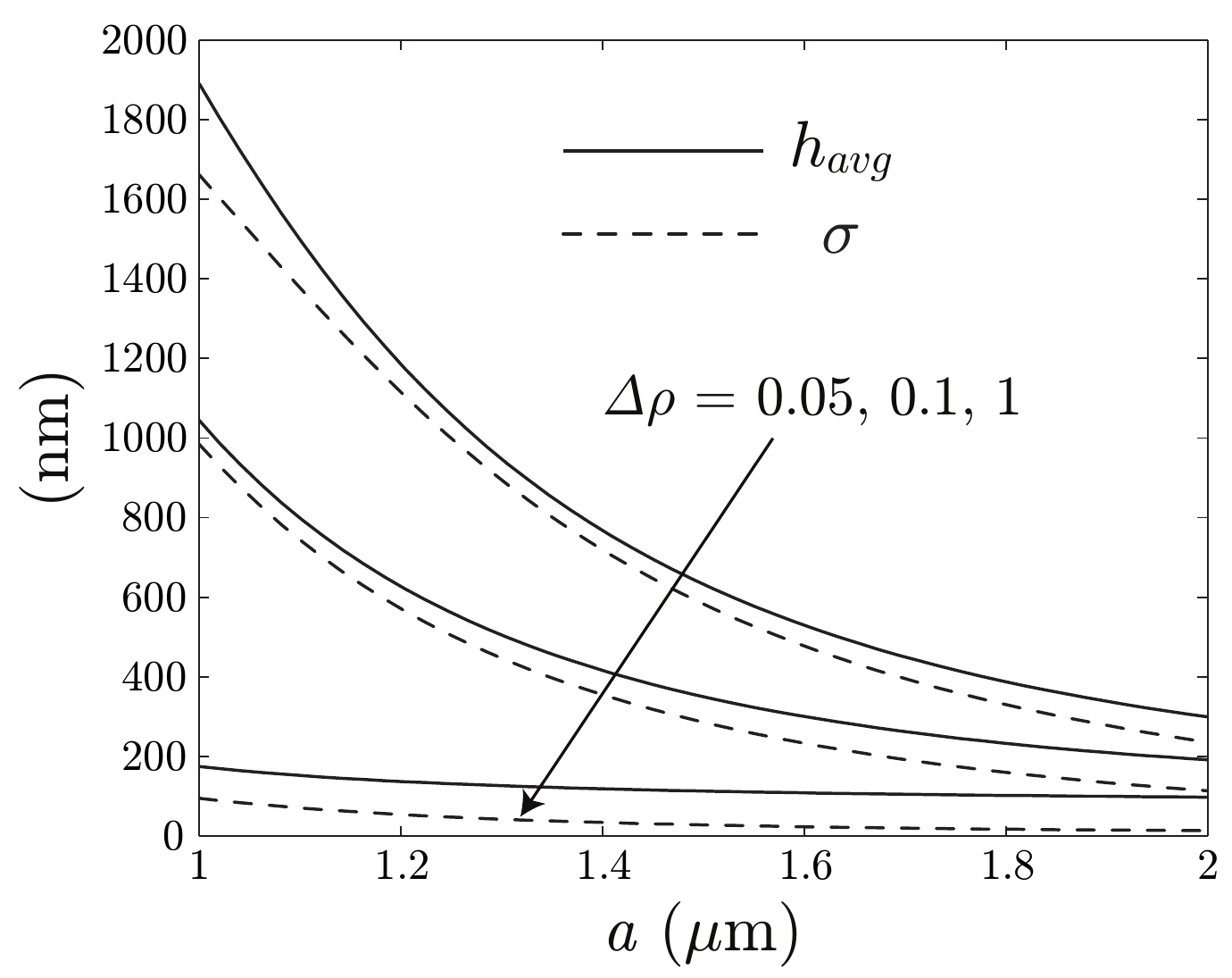}}  \\ 
			 \hline				\end{tabular}
	\end{center}
\caption[Average separation]{Average particle separation $-$solid curve$-$ and corresponding standard deviation $-$dashed curve$-$. The standard deviation is comparable to the average separation for  the particles in the low range of buoyant weights considered. As the buoyant density and the size of the particle increase, the particles are closer to the wall and more narrowly confined. The parameters of the electrostatic and Van der Waals potentials are as noted in the caption to Fig.~\ref{Fig:potentials}. The arrow points in the direction of increasing buoyant density.} 
\label{Fig:avg_separation}
\end{figure}

The Van der Waals potential can be rigorously computed via Lifshitz theory and fit for convenience to a power law expression of the form~\citep{Wu:2006}
\begin{equation}
V_{vdw}(h) = -a A h^{-p}.
\label{VdWpotential}
\end{equation}
These surface potentials have been referenced to zero at an infinite separation. The gravitational potential is given in terms of the buoyant weight of the particle
\begin{equation}
V_{grav}(h) = \left.4/3\pi a^3 \Delta \rho g \left(h+h_{ref}\right)\right/\bc T,
\end{equation}
where $\Delta \rho  = \rho_p - \rho_m$ is the density difference between the particle and the medium $-$or buoyant density$-$, $g$ is the acceleration due to gravity, and $h_{ref}$ is the height of channel wall with respect to an arbitrary reference. Fig.~\ref{Fig:potentials} shows the total potential for particles with a radius of 2 $\mu$m as a function of the particle-wall separation, calculated using parameters representative of the experimental values reported by Wu et al~\citep{Wu:2006} for two values of the buoyant density. Specifically, Stern potentials of -60 mV, and Van der Waals afinity prefactor and exponent of of 2 nm and 2, respectively, are representative of the experimental values corresponding to a silica particle suspended above a bare glass surface in a 1 mM aqueous solution of a 1:1 electrolyte. In this case the surface interactions  are important within a few hundred nanometers from the wall with gravity dominating for larger separations. The less dense particle is confined to a broader range of separations and is, on the average, farther away from the wall than the denser particle $-$see panels $(a)$ and $(b)$ in Fig.~\ref{Fig:potentials}, respectively. Fig.~\ref{Fig:avg_separation} shows the average particle-wall separation and the standard deviation as a function of particle radius for different buoyant densities. The particles are on average closer to the wall and more narrowly confined as the particle buoyant density and size increase, with the excursions of the heavier particles restricted within tens of nanometers from their equilibrium position. This degree of confinement makes the FJ approximation valid and in fact accurate even at relatively large P\'eclet numbers, as shall be shown later by comparison with Brownian dynamic simulations.

\subsection{Inducing partition via physical and chemical patterns}

In the absence of a driving field, the concentration of particles undergoing Brownian motion over a stripe pattern as shown in Fig.~\ref{Fig:Schematic} follows the Boltzmann distribution. In this case, the marginal distribution is given by
\begin{equation}
{\cal{P}}(y) = \left\{\begin{array}{c} \frac{1}{\epsilon + (1-\epsilon)K}   \quad     0    < y < \epsilon \\
                                                   \frac{K}{\epsilon + (1-\epsilon)K}   \quad \epsilon < y < 1 .
                           \end{array}
                    \right.,
\end{equation}
with the partition ratio, $K$, defined as
\begin{equation}
K = \left. \int e^{-V_{2}(z)}dz \right/\int e^{-V_{1}(z) }dz.
\label{Eq:K}
\end{equation}
It is clear that differences in the potential energy of the particle in the two regions would result in partition, $K$ being a measure of the relative affinity between the particle and each of the stripes. 

\begin{figure}[htbp]
	\begin{center}
		\begin{tabular}{|c|c|}
			\hline
			  $(a)$  & $(b)$ \\
			 \parbox[][][c]{6cm}{
	          		\includegraphics[width=6cm]{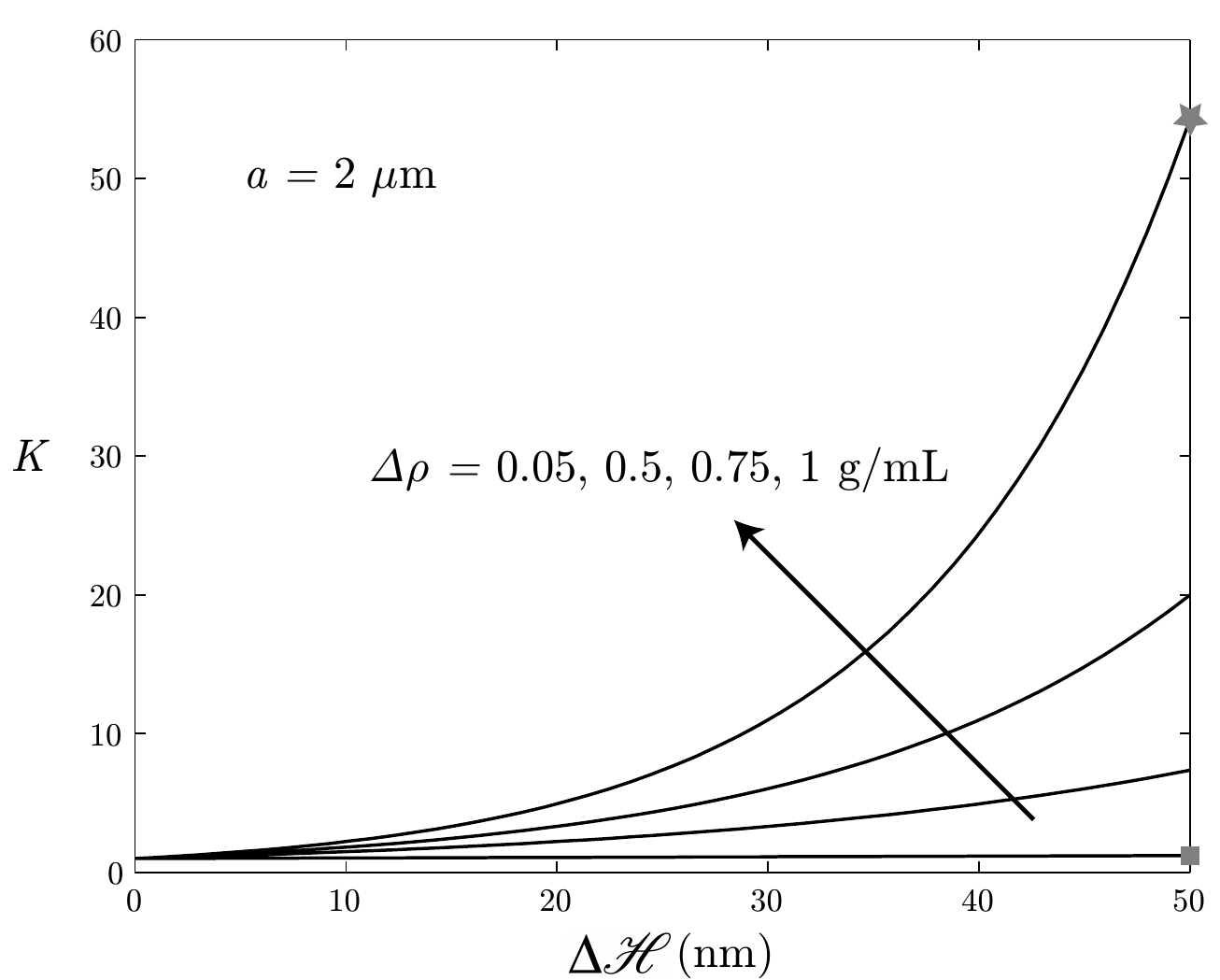}} &
			 \parbox[][][c]{6cm}{
	          		\includegraphics[width=6cm]{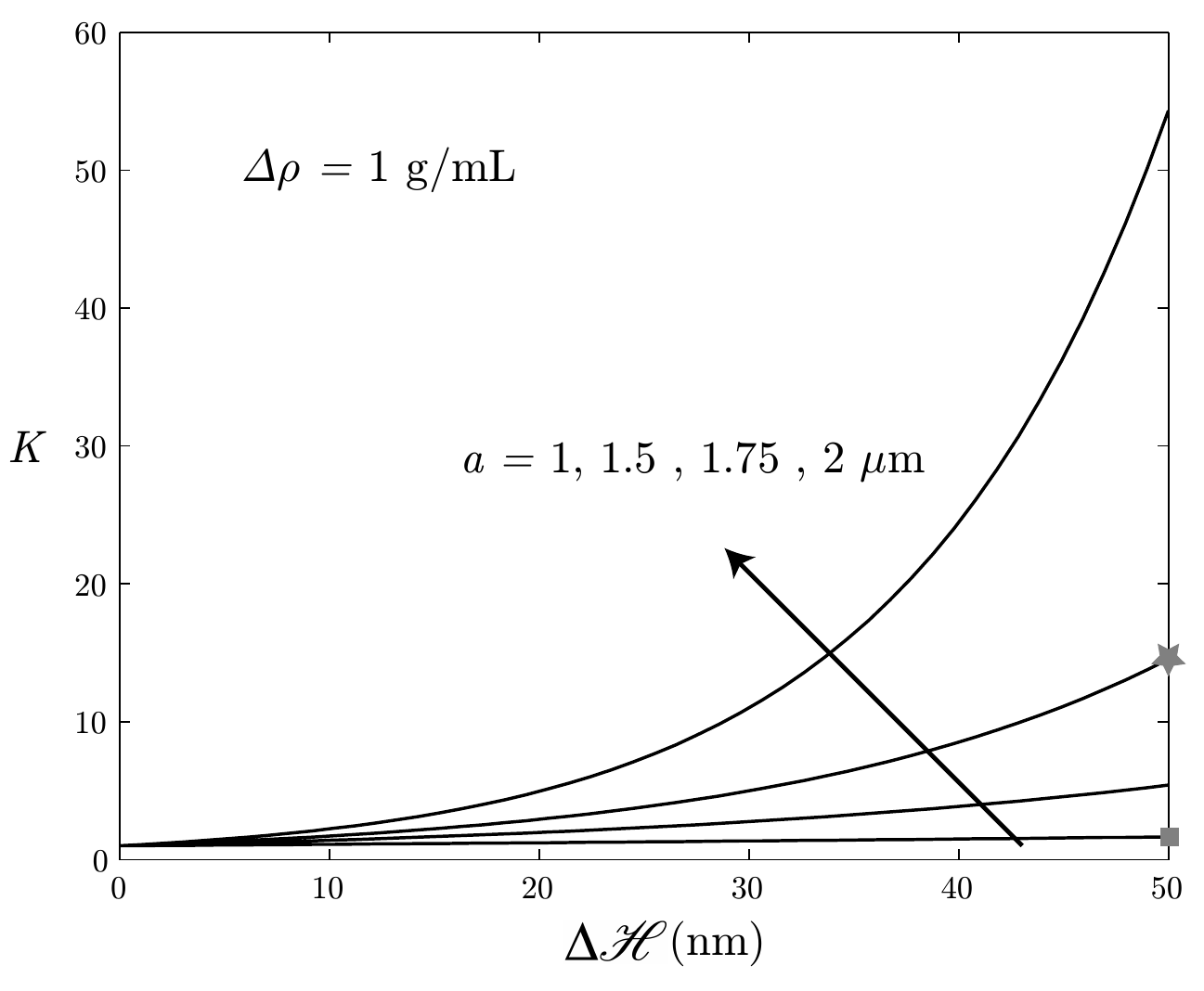}} \\
	     \hline
	      $(c)$ &  $(d)$ \\    		
	     \parbox[][][c]{6cm}{
	          		\includegraphics[width=6cm]{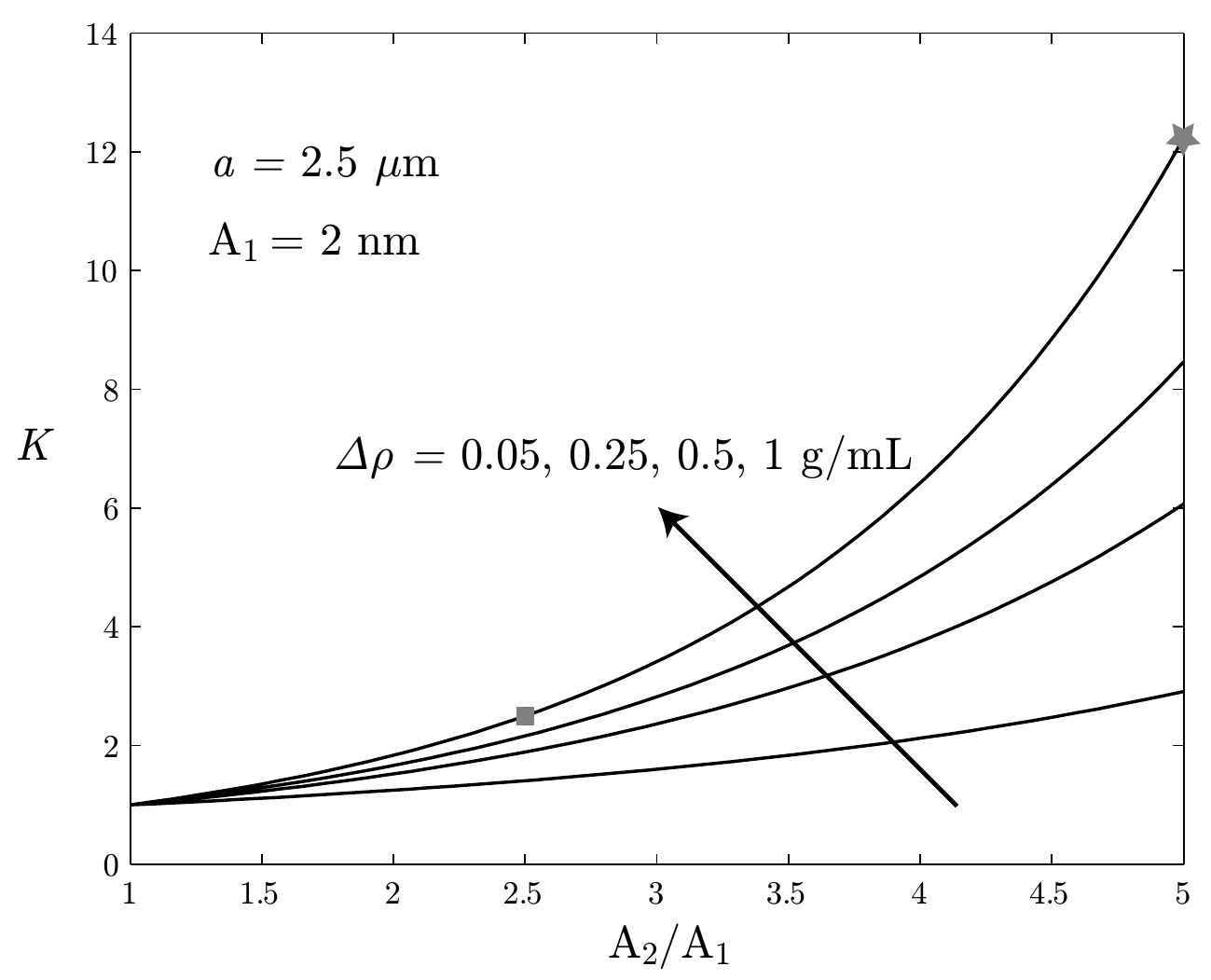}}    &
	     \parbox[][][c]{6cm}{
	          		\includegraphics[width=6cm]{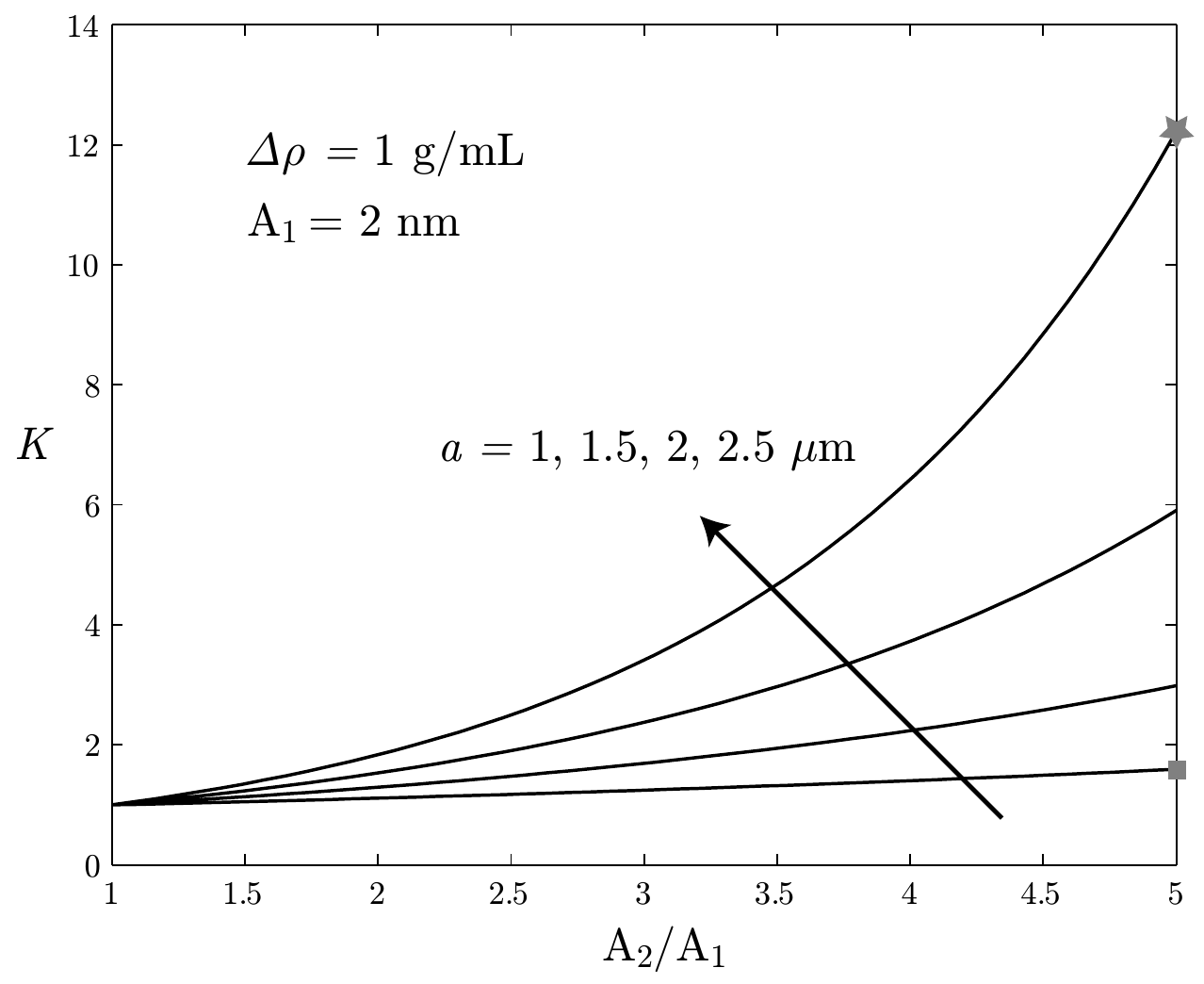}} \\
			\hline			
		\end{tabular}
	\end{center}
\caption[Partition coefficient]{Partition ratio induced $-$$(a)$ and $(b)$$-$ by gravity in a pattern of chemically identical stripes as a function of the height difference between the two stripes, and  $-$$(c)$ and $(d)$$-$ by Van der Waals interactions as a function of the particle-surface affinity with one of two otherwise identical stripes. In $(a)$ and $(c)$ the partition ratio is plotted for different values of the buoyant density, whereas in $(b)$ and $(d)$ the partition ration is plotted for different particle sizes. The arrow traverses curves of increasing value of the respective parameter. For all cases, the Stern potential of the particle and stripes is -60 mV, the molarity of the aqueous solution $-$1:1 electrolyte$-$ is 1 mM , and the Van der Walls exponent is $p = 2$.}
\label{Fig:K}
\end{figure}

Patterning the stripes chemically would result in different surface potentials while patterning the stripes physically modifying the topography of the surface would result in different sampling of the gravitational potential. Fig.~\ref{Fig:K} shows partition ratios predicted from Eq.~(\ref{Eq:K}) with the electrostatic and Van der Waals  potentials based on the experimental values reported by Wu et al~\citep{Wu:2006}. Specifically, we first show the partition ratio induced by gravity in a pattern of chemically identical stripes as a function of the height difference between them for $-$Fig.~\ref{Fig:K}$(a)$$-$ different values of the buoyant density, ranging approximately from that of polystyrene particles to that 
of silica particles, and $-$Fig.~\ref{Fig:K}$(b)$$-$ for different values of the particle radius and for a buoyant density representative of a silica particle. In this case, in which the surface properties of both stripes are the same, Eq.~(\ref{Eq:K}) for the partition ratio simplifies to
\begin{equation}
K = e^{\left.4/3\pi a^3 \Delta {\cal{H}} \Delta \rho g\right/\bc T},
\end{equation}
where $\Delta {\cal{H}}$ is the height difference between the stripes. The partition ratio increases with the height difference between the stripes, the buoyant density and the size of the particle. It is worth mentioning that the sampling of the transverse direction is such that the average particle-wall separation is larger than the largest $\Delta {\cal{H}}$ considered $-$see Fig.~\ref{Fig:avg_separation}$-$, and therefore the particles do not experience physical hindrance at the transition regions. In panels $(c)$ and $(d)$ of Fig.~\ref{Fig:K} we present partition induced via Van der Waals interactions by increasing the particle affinity, parameter $A$ in Eq.~(\ref{VdWpotential}), with one of two $-$otherwise identical$-$ stripes. Specifically, a value of $p$ of 2 and values of $A$ between 2 and 10 nm are representative of the range spanned in the case of a silica particle in a 1 mM aqueous solution of a 1:1 electrolyte interacting with a glass substrate coated with a gold film of thickness between 0 and 30 nm. In this case the Van der Waals-induced partition ratio increases as the affinity ratio, the buoyant density, or size of the particle increase.
	
\subsection{Non-equilibrium transport}
\begin{figure}[htbp]
	\begin{center}
		\begin{tabular}{|c|c|c|}
			\hline
			  $(a)$  & $(b)$ \\
			 \parbox[][][c]{6cm}{
	          		\includegraphics[width=6cm]{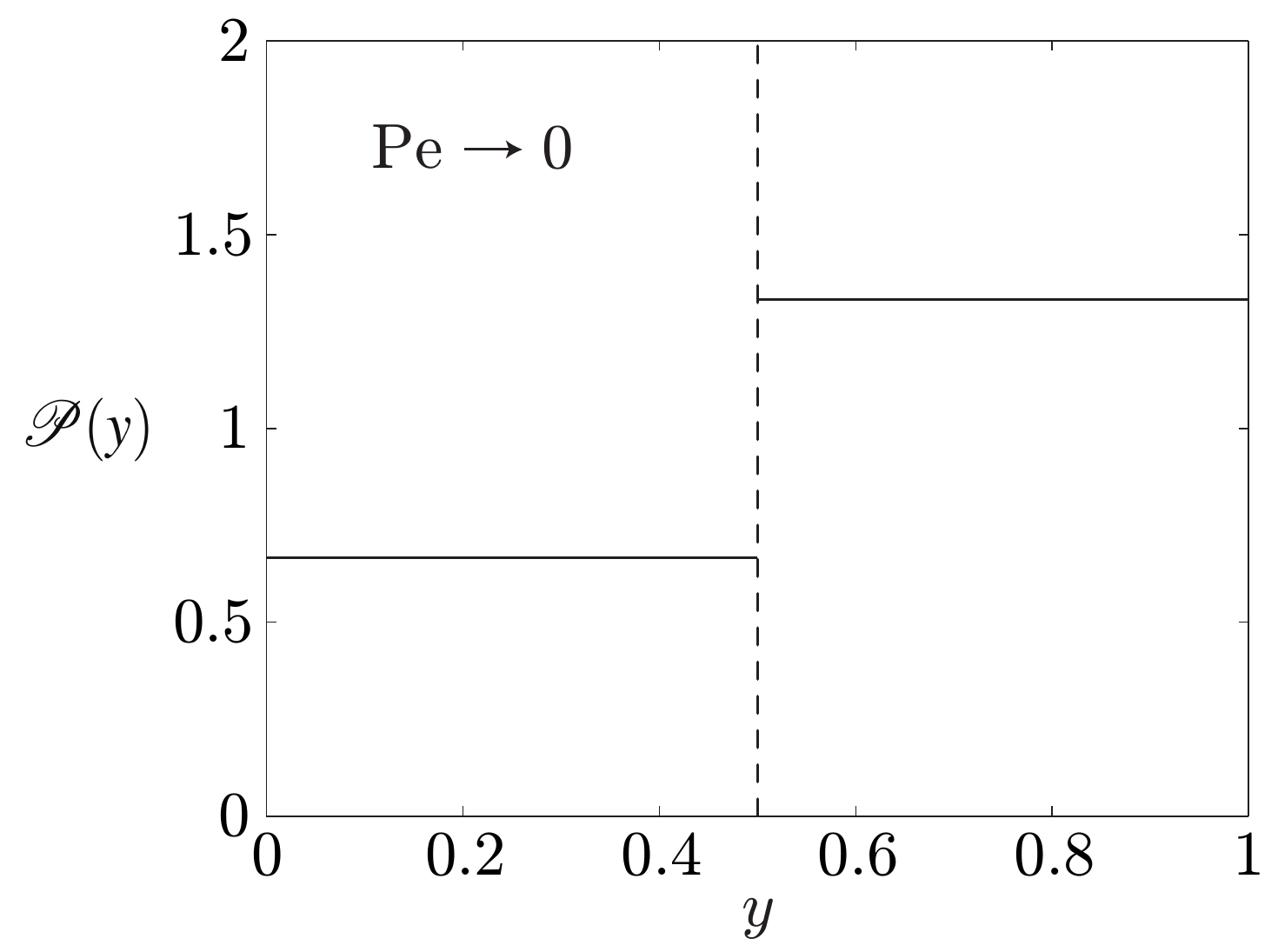}} & 
			 \parbox[][][c]{6cm}{
	          		\includegraphics[width=6cm]{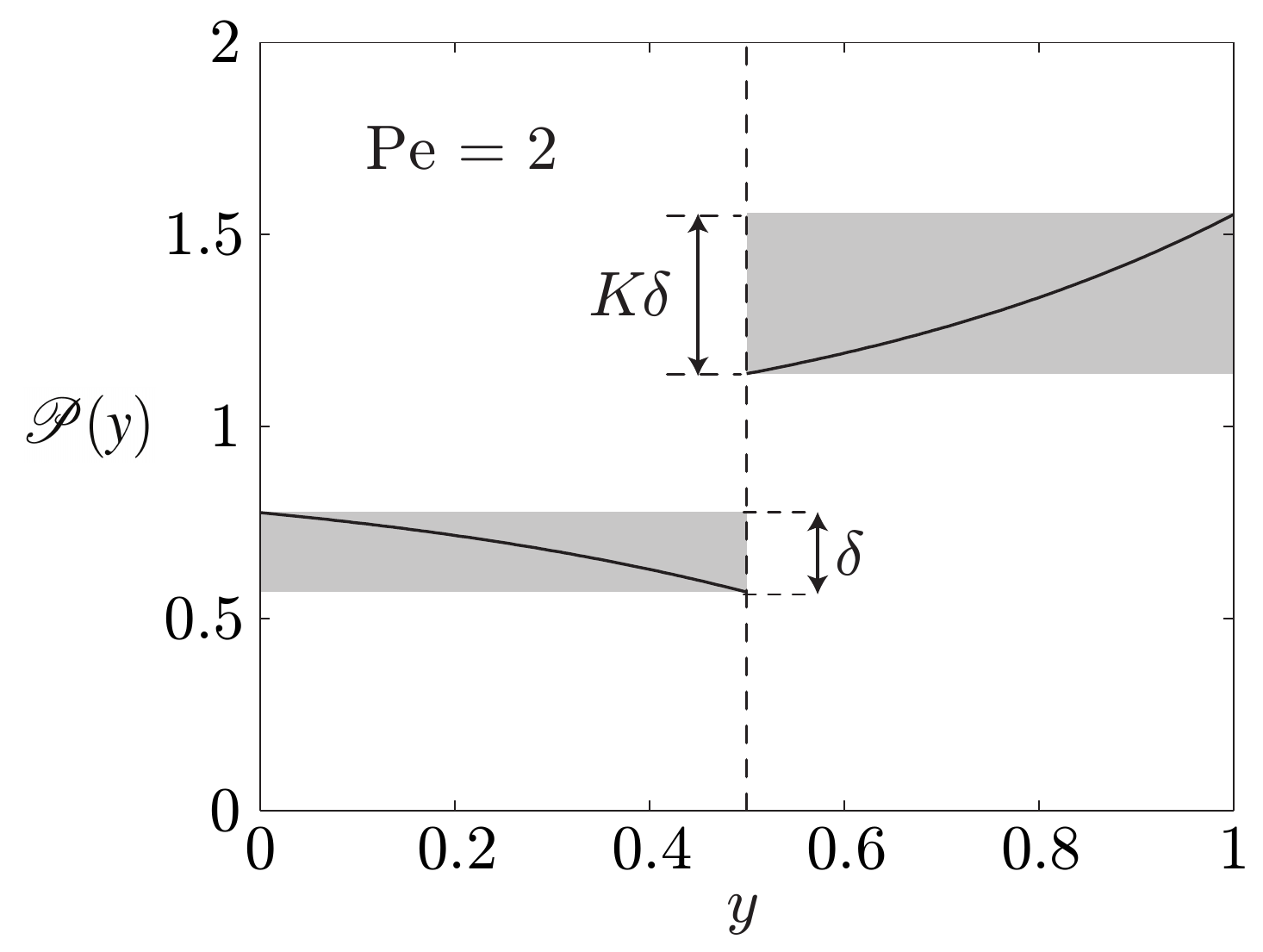}}   \\
	     \hline
	      $(c)$ &  $(d)$ \\    		
	     \parbox[][][c]{6cm}{
	          		\includegraphics[width=6cm]{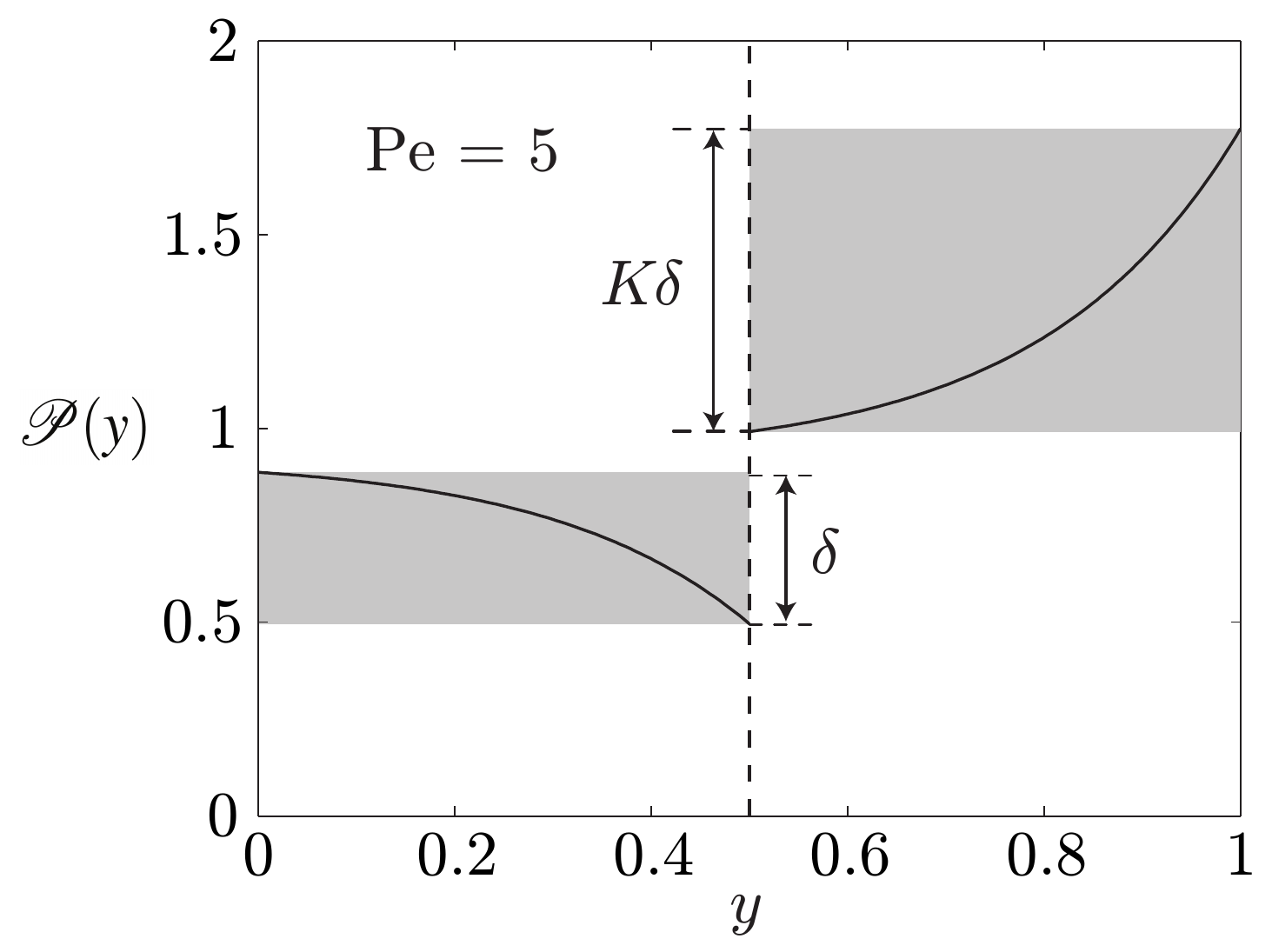}}    &
	     \parbox[][][c]{6cm}{  
	          		\includegraphics[width=6cm]{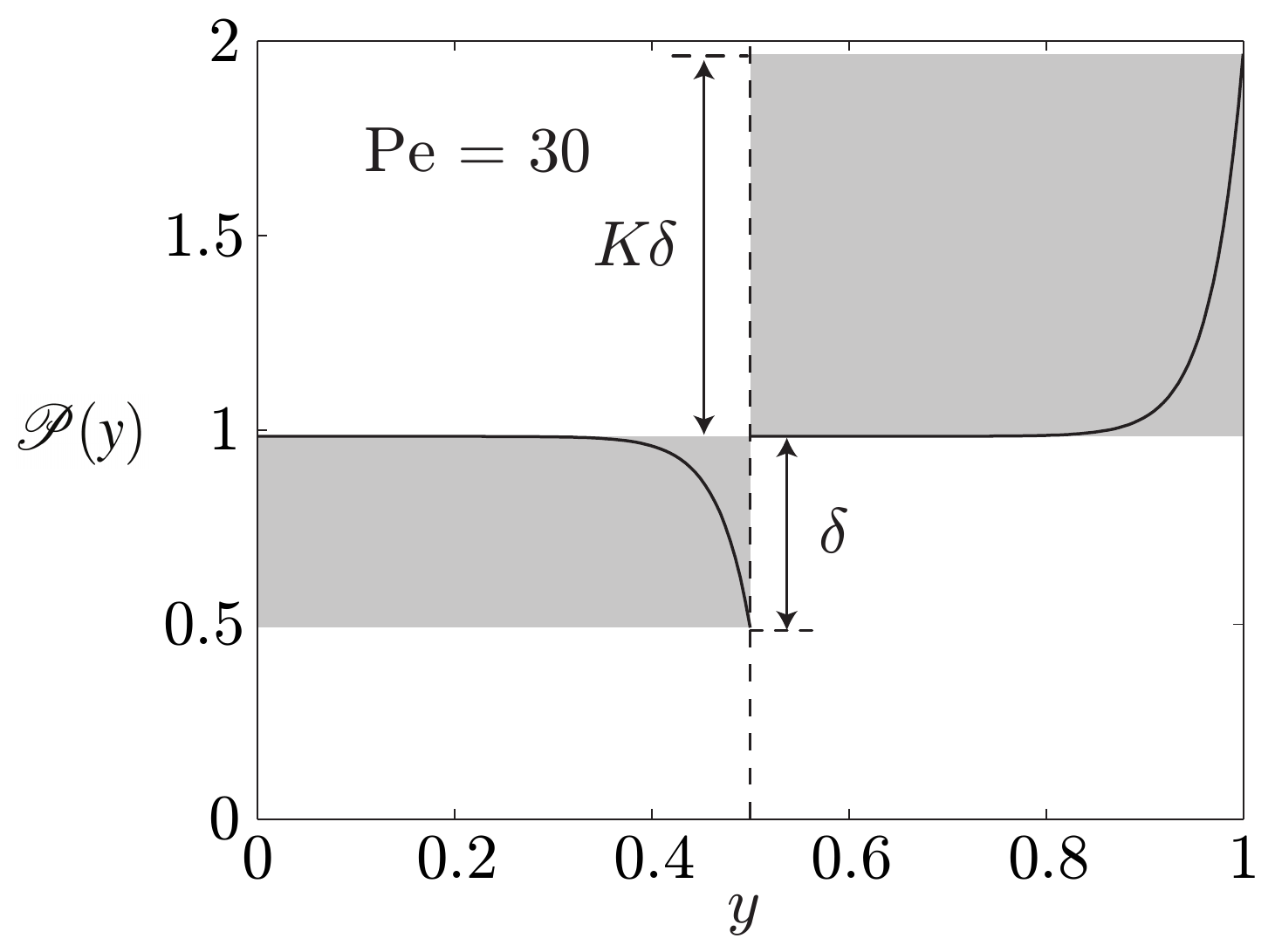}}   \\
			\hline			
		\end{tabular}
	\end{center}
\caption[Marginal probability density]{Marginal probability density for different P\'eclet numbers for a partition ratio $K = 2$. The differences in the convective fluxes at the interface between the stripes due to partition result in compensating diffusive fluxes to satisfy continuity of the flux. The shaded areas represent the concentration difference across the stripes, which differ by a factor of $K$. The net diffusive flux opposes the total flux in the direction normal to the stripes.}
\label{Fig:mpd}
\end{figure}

We now consider the cases in which the particle is either driven by a spatially uniform external force {\mathbi{\itshape{F}}} oriented at an angle $\theta_{\mbox{\itshape{\tiny{F}}}}$ with respect to the $y$ axis $-$see Fig.~\ref{Fig:Schematic}$-$, or convected by a flow field that far from the particle is oriented at an angle $\theta_{f}$ with respect to the $y$ axis. The migration of particles at angles that differ from the forcing direction can be explained qualitatively by considering the normal flux at the transition between the stripes. Consider the equilibrium marginal probability density shown in Fig.~\ref{Fig:mpd}$(a)$ for a partition ratio $K =2$. In the presence of a driving field, the convective flux on each side of the interfaces between the stripes would differ due to the concentration differences. Thus, for the system to reach steady state and the corresponding continuity of the total flux, concentration gradients and the ensuing diffusive fluxes must compensate for the differences in the convective fluxes. As can be seen in Fig.~\ref{Fig:mpd} $(b)$-$(d)$, the diffusive fluxes augment $-$reduce$-$ the flux in the region of lower $-$higher$-$ concentration and convective transport. The average diffusive flux on each stripe is proportional to the concentration difference across them $-$height of the shaded regions shown in Fig.~\ref{Fig:mpd}$-$. We also show in Fig.~\ref{Fig:mpd} that the average diffusive fluxes differ by a factor $K$. The net result is a diffusive component of the flux normal to the stripes that reduces the total flux in this direction, the effect being stronger for increasing partition ratios. As a consequence, the trajectory angle of the particle is different from the orientation angle of the field and strongly depends on the partition ratio. This is the basis for the separation of particles exhibiting different partition ratios. 

Evaluating the general expression for the chromatographic trajectory angle, Eq.~(\ref{Eq:theta_F_arbitrary}), for the piece-wise potential given by Eq.~(\ref{Eq:piece_wise_potential}), we obtain the trajectory angle in terms of the partition ratio for the case in which the particles are driven by an external force
\begin{equation}
\tan\theta^* =  \tan \theta_F \left\{ 1 + \frac{\left(1- \lambda K\right)^2}{\lambda K}\frac{\left(1-e^{-\tinyPe \epsilon}\right)\left[1-e^{-\tinyPe(1- \epsilon)}\right]}{1-e^{-\tinyPe}}\frac{1}{\Pe} \right\},
\label{Eq:theta_F_stripes}
\end{equation}
where $\lambda = \bar{M}_{||_2}/\bar{M}_{||_1}$ with $\bar{M}_{||_{\mbox{\scriptsize{i}}}}$ the local average of the mobility over stripe i defined by Eq.~(\ref{Mbar}). In the limit of vanishing $\Pe$ the above expression reduces to
\begin{equation}
\tan\theta^* =  \tan \theta_F \left[ 1 + \frac{\left(1- \lambda K\right)^2}{\lambda K}\epsilon(1-\epsilon)\right], 
\label{Eq:theta_F_stripes_Pe_to_0}
\end{equation}
while in the convection dominated case the trajectory angle tends to the forcing angle. 

Analogously, replacing the piece-wise potential into Eq.~(\ref{Eq:theta_Flow_arbitrary}) yields the migration angle for the case in which the particle is convected by a fluid flow 
\begin{equation}
\tan\theta^* =  \tan \theta_f \left\{ 1 + \frac{\left(1- \lambda K\right)(1-\Pi \lambda K)}{\Pi \lambda K}\frac{\left(1-e^{-\tinyPe \phi_1 \epsilon}\right)\left[1-e^{-\tinyPe \phi_1 \Pi(1- \epsilon)}\right]}{1-e^{-\tinyPe \phi_1[\epsilon+\Pi(1-\epsilon)]}}\frac{1}{\Pe \phi_1} \right\} 
\label{Eq:theta_Flow_stripes}
\end{equation}
where $\Pi = \phi_2/\phi_1$, with $\phi_{\mbox{\scriptsize{i}}} = \bar{u}_{\mbox{\scriptsize{i}}} /\bar{M}_{||_{\mbox{\scriptsize{i}}}}$ being the effective  hydrodynamic force acting on the particles moving on stripe i. 

In the limit as the $\Pe$ tends to zero the above expression simplifies to
\begin{equation}
\tan\theta^* =  \tan \theta_f \left[ 1 + \frac{\left(1- \lambda K\right)(1-\Pi \lambda K)}{\lambda K}\frac{\epsilon(1-\epsilon)\Pi}{\epsilon+\Pi(1-\epsilon)}\right], 
\end{equation}
and as in the external force case, the migration angle converges to the forcing angle in the high $\Pe$ number limit.

\subsection{Brownian dynamic simulations}
The expressions for the migration angle in the case of stripe patterns, Eqs.~(\ref{Eq:theta_F_stripes}) and~(\ref{Eq:theta_Flow_stripes}), are valid under the FJ approximation, i.e., under the assumption of instantaneous equilibrium in the transverse direction. This assumption is clearly valid when the particle has ample time to equilibrate in the direction perpendicular to the  patterned surface via diffusion as it is convected across the stripes by either an external force or by a fluid flow. Therefore, the relevant parameter is the ratio between the diffusive time in the transverse direction and the convective time along a single cell, i.e., $\Pe^{FJ} = \Pe \left(\sigma/l_y\right)^2$ where $\sigma$ is the width of the confinement region. Then, if the particles are narrowly confined to the vicinity of the patterned surface the FJ approximation could be valid even for large values of the P\'eclet number~\citep{Burada:2007,XinLi:2009}. The particles could be physically constrained to move in a narrow channel or, alternatively, a highly confining potential could restrict the vertical motion of the particles to a narrow region independently of the height of the channel. Therefore, in order to determine the full range of validity of the results presented in the preceding section and, more important, the range of $\Pe$ that result in separation, Brownian dynamic $-$BD$-$ simulations were carried out for the case when the particle is convected by a fluid flow. Using the Ermak and McCammon algorithm~\citep{Ermak_McCammon:1978}, the dimensionless equations of motion for a suspended particle are
\begin{subequations}
\begin{equation}
\Delta x = \xi  \Pe \upsilon^0 \sin \theta_f\Delta t + \xi \sqrt{2 \lambda^0_{||} \Delta t} \enspace \chi 
\end{equation}
\begin{equation}
\Delta y = \Pe \upsilon^0 \cos \theta_f \Delta t - \lambda^0_{||}\left(\frac{\partial V}{\partial y} \right)^0 \Delta t + \sqrt{2 \lambda^0_{||} \Delta t} \enspace \psi 
\end{equation}
\begin{equation}
\Delta z = \gamma^2 \left(\frac{d\lambda_{\perp}}{dz}\right)^0 \Delta t - \gamma^2 \lambda^0_{\perp}\left(\frac{\partial V}{\partial z} \right)^0  \Delta t + \gamma \sqrt{2 \lambda^0_{\perp} \Delta t} \enspace \omega 
\end{equation}
\end{subequations}
The characteristic scales used to nondimensionalize the equations are $l_x$, $l_y$, and $d$ in the corresponding directions, the unbounded mobility $-M_{\infty}-$ and diffusivity $-D_{\infty}-$ for the corresponding quantities, the average velocity of the particle-free flow $-U-$ for the particle velocity, $k_B T$ for the interaction potential, and $l^2_y/D_{\infty}$ for the time step. In the above equations $\xi = lx/ly$, $\gamma = lz/ly$, $\upsilon$ is the dimensionless particle velocity $-$in the forcing direction$-$ due to the flow, $\lambda$ is the dimensionless mobility $-$or diffusity$-$, $\Pe = U l_y /D_{\infty}$, $\theta_f$ is the orientation angle of the particle free flow, and $\chi$, $\psi$, and $\omega$ are independent random numbers chosen from a symmetric distribution of zero mean and unit width. Lastly, the superscript indicates that the quantities are evaluated at the beginning of the time step.

In order to carry out the simulations it is necessary to specify transition potentials. As the particle traverses the transition region from one stripe to the other, it experiences a force toward the stripe of higher affinity. The magnitude of this force along the transition region is dictated by the imbalance in the particle-substrate affinity between the anterior and posterior regions of the particle. The exact form of the transition potential would depend on physico-chemical properties of the transition and these in turn would depend on the specific way in which the pattern was created. Exact expressions for electrostatic and Van der Waals interaction potentials are not available even for the case in which the properties of a half-space change discontinously at a point resulting in two regions of homogeneous but different properties. We model the transition potential as an s-shape curve given by a third degree polynomial in $y$ constructed to satisfy continuity of the total potential and yield a vanishing force at the edges of the transition regions at every height. 


\subsection{PIVC in microfluidic devices}
\begin{figure}[htbp]
	\begin{center}
		\begin{tabular}{|c|c|}
			\hline
			  $(a)$  & $(b)$ \\
			 \parbox[][][c]{6cm}{
	          		\includegraphics[width=6cm]{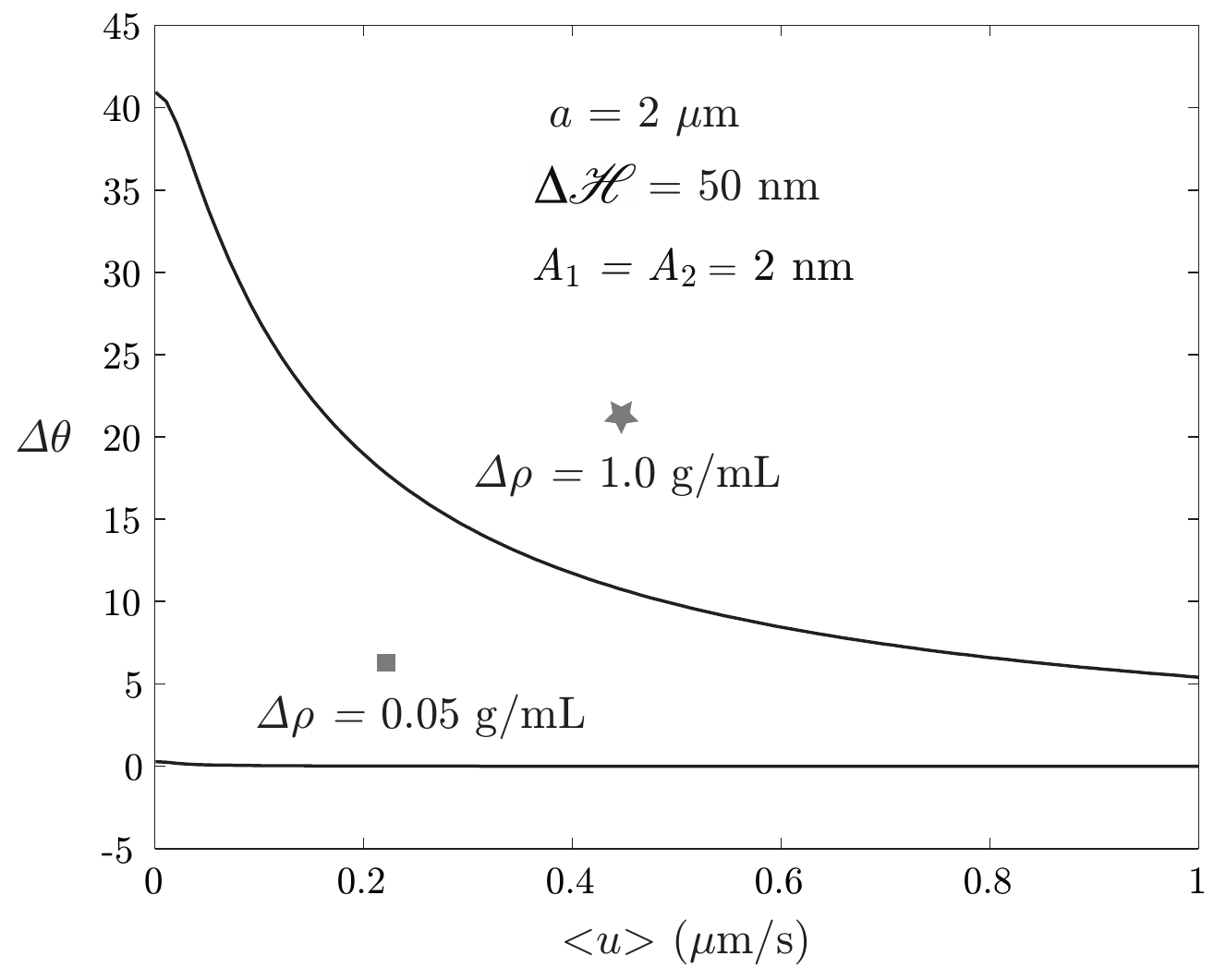}} &
			 \parbox[][][c]{6cm}{
	          		\includegraphics[width=6cm]{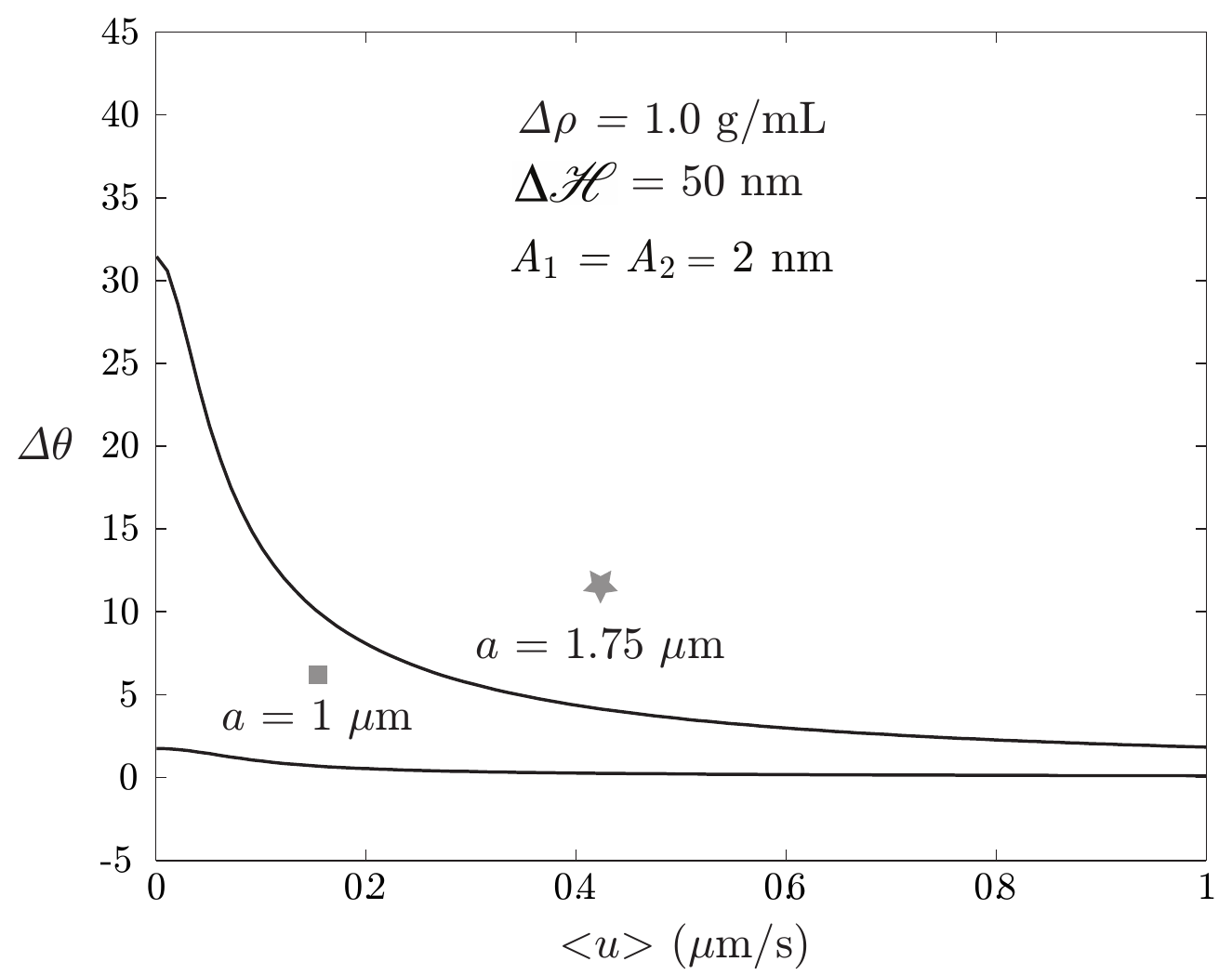}} \\
	     \hline
	      $(c)$ &  $(d)$ \\    		
	     \parbox[][][c]{6cm}{
	          		\includegraphics[width=6cm]{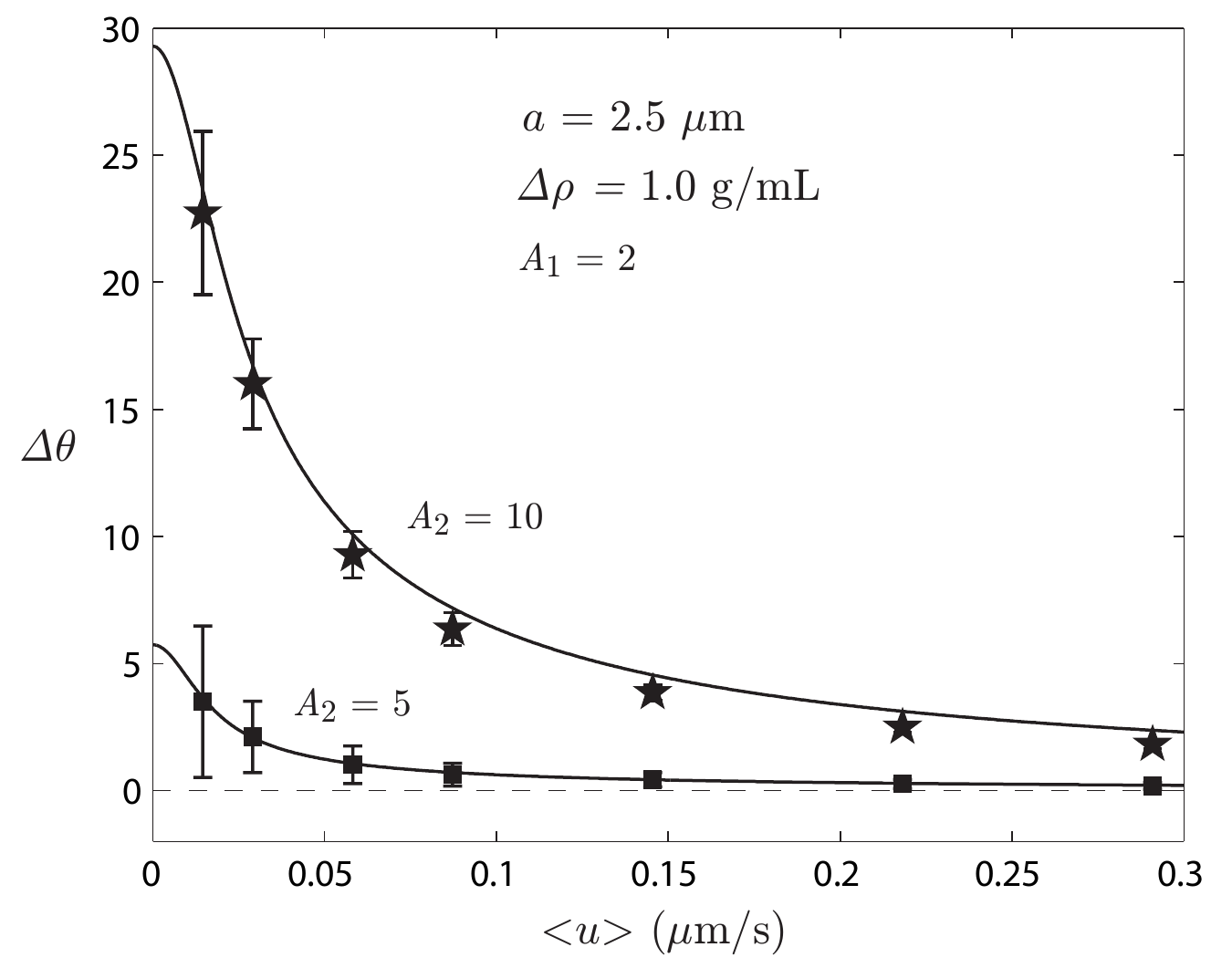}}    &
	     \parbox[][][c]{6cm}{
	          		\includegraphics[width=6cm]{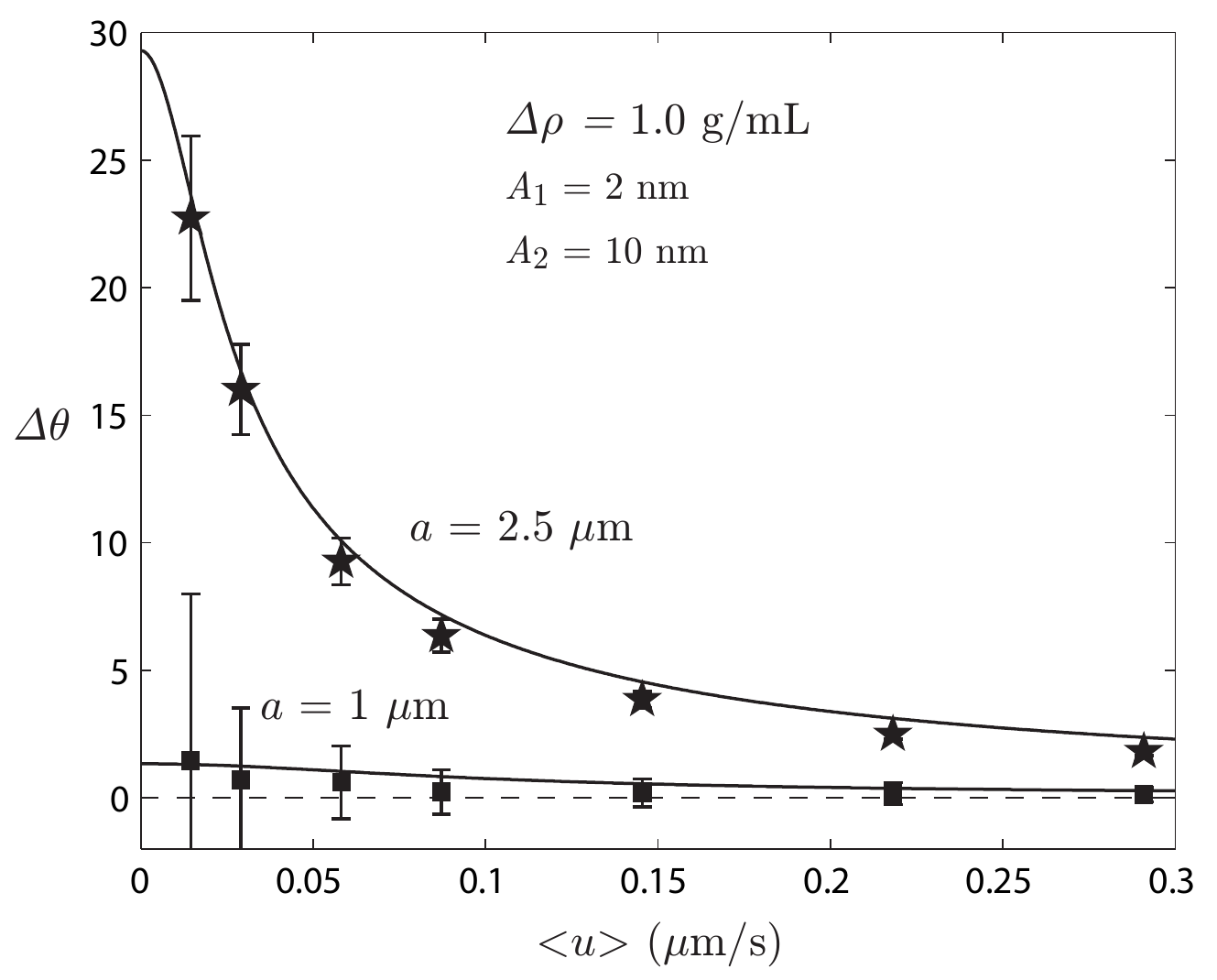}} \\
			\hline			
		\end{tabular}
	\end{center}
\caption[Discrimination angles as a function of the average velocity of the underlying flow with partition induced by gravity and Van der Waals interactions for particles of the same and different size]{Discrimination angles as a function of the average velocity of the underlying flow with partition induced by gravity, $(a)$ and $(b)$,  and Van der Waals interactions, $(c)$ and $(d)$, for particles of the same and different sizes. The corresponding partition ratios are indicated by the solid square and star in the plots in Fig.~\ref{Fig:K} with the same labels.}
\label{Fig:theta}
\end{figure}

To show the potential of PIVC we consider different separation scenarios of a mixture of particles carried by a pressure-driven flow through the stripe pattern shown in Fig.~\ref{Fig:Schematic} and compare the predictions based on the FJ approximation with the results of BD simulations. The geometry considered is a slit channel with a height of 10 $\mu$m. The bottom wall is patterned with 10 $\mu$m stripes, oriented at a forcing angle of $45$ degrees with respect to the average flow $-\theta_f = 45^o-$.  To predict the migration angles from Eq.~(\ref{Eq:theta_Flow_stripes}) we use the  analytical approximation for the particle mobility given by Pawar and Anderson~\citep{Pawar:1993} and interpolate the numerical results tabulated by Staben et al.~\citep{Staben:2003} for the velocity of a particle confined between two parallel walls. Let us note that in the case of physical patterns we neglect the effect of the steps in the hydrodynamic force and resulting velocity of the particle. Fig.~\ref{Fig:theta} presents the discrimination angle, i.e., the difference between the trajectory angle and the forcing angle, for different pairs of particles as a function of the average velocity of the carrying flow.  Specifically, each of the systems and pair of particles considered in panels $(a)$-$(d)$ in Fig.~\ref{Fig:theta} correspond to a system and particles for which the partition ratio is shown in panels $(a)$-$(d)$ in Fig.~\ref{Fig:K}, respectively. The correspondence is indicated by solid squares and stars. In Fig.~\ref{Fig:theta}$(a)$ $-(b)-$ particles of the same $-$different$-$ size are separated according to their mass with partition induced by gravity. Panels $(c)$ and $(d)$ in Fig.~\ref{Fig:theta} show the separation of particles based on their Van der Waals interaction with chemically different stripes. Fig.~\ref{Fig:theta}$(c)$ shows the separation of particles of the same size but with different affinity with one of the stripes. The separation of particles of the same material but of different sizes is shown in Fig.~\ref{Fig:theta}$(d)$. Separation is readily accomplished for all cases considered; one of the particles only marginally departs from the forcing angle owing to weak partition, while the other particle is restrained to one of the stripes resulting in strong deviations from the forcing angle. The data points and respective error bars correspond to the average and standard deviation, respectively, of 128 realizations of BD simulations tracking the particle for 50 units of dimensionless time and with a transition region with a width of two-hundredth the length of the unit cell in the $y$-direction. The FJ approximation and the simulation results are in excellent agreement over the whole range of discrimination angles due to the narrow confinement of the particles by the potential $-$see Fig.~\ref{Fig:avg_separation}$-$.

\section{Conclusions}

We showed that vector chromatography in planar microfluidic devices is feasible by harnessing surface interactions that lead to the spontaneous partition of different species. First, we derived analytical solutions for the trajectory angle of a particle driven by an external force or convected by a flow field in a slit geometry for an arbitrary two-dimensional periodic potential under the Fick-Jacobs approximation.  We showed that the migration angle depends on particle properties, thus providing the basis for vector chromatography of different species. We considered the case of a piece-wise constant periodic potential that could be created by chemically or physically patterning one of the surfaces of a microfluidic device with an array of rectangular stripes, thus causing the spontaneous partition of suspended particles. We showed that partition results in diffusive fluxes that oppose the total flux and reduce the velocity component in the direction normal to the stripes making the particles migrate at angles that differ from the orientation angle of the driving field, the effect being stronger with increasing partition ratios. Specifically, we considered the fractionation of particles of the same and different size using both physically patterned channels where partition is induced by 1-g gravity as well as chemically patterned channels in which partition is induced by Van der Waals interactions. We validated our results by means of BD simulations that agree well with the results obtained by means of the FJ approximations due to the highly confining nature of the particle-wall interaction.  Thus, we demonstrated that partition-induced vector chromatography $-$PIVC$-$ of particles exhibiting different partition ratios is feasible in microfluidic devices. In addition to  continuous operation, another promising feature of PIVC is its potential versatility that stems from the vast range of physicochemical properties that can be employed to induce partition. Furthermore, separation can be achieved without the need for external components making the device autonomous, thus facilitating its portability and integration with other lab-on-a-chip components. In addition, partition can also be induced or enhanced by means of externally applied fields. As in FFF, different subtechniques are envisioned depending on the nature of the forces use to cause partition. Further work is needed to explore different pattern geometries besides the stripe case considered here in order to optimize the sensitivity of the discrimination angle to the partition ratio and extend the range of operation to higher P\'eclet numbers.  Lastly, it is interesting to note that the results presented here suggest the possibility of inducing vector chromatography by exploiting the dependence of the trajectory angle on the velocity and mobility of the particles over each stripe through the parameters $\lambda$ and $\Pi$. 

\bibliographystyle{model1a-num-names}
\bibliography{bibliography}







\end{document}